\documentclass[12pt]{JHEP3}
\usepackage{epsfig}
\preprint{ {\tt{ hep-th/0502186}}\\ {  SPIN-05/06} \\ {ITP-05/08}}
\newcommand{\be}[1]{ \begin{equation}\label{#1} }
\newcommand{\ee}{\end{equation}}
\newcommand{\bea}[1]{\begin{eqnarray}\label{#1} }
\newcommand{\eea}{\end{eqnarray}}
\newcommand{\eq}[1]{(\ref{#1})}
\newcommand{\fig}[1]{fig.\,\ref{#1}}
\newcommand{\del}{\partial}

\title{  Structure constants of planar
${\cal N} =4$ Yang Mills at one loop }
\author{ Luis F. Alday$^{a}$, 
Justin R. David$^b$, Edi Gava$^{b,c}$ , K. S. Narain$^b$ \\
$^a$Institute for Theoretical Physics and Spinoza Institute, \\
Utrecht University, 3508 TD Utrecht, \\
The Netherlands. \\
$^b$High Energy Section, \\
The Abdus Salam International Centre for Theoretical Physics,
\\Strada Costiera, 11-34014 Trieste, Italy.\\
$^c$Instituto Nazionale di Fisica Nucleare, sez. di Trieste, \\
and SISSA, Italy. \\
\email{L.F.Alday@phys.uu.nl, 
justin, gava, narain@ictp.trieste.it} 
}
\abstract{ 
We study structure constants of 
gauge invariant operators in planar ${\cal N}=4$ Yang-Mills 
at one loop with the motivation of determining features of 
the string dual of weak coupling Yang-Mills. 
We derive a simple  
renormalization group invariant formula
characterizing the corrections to structure constants of any
primary operator in the planar limit. Applying this to the scalar
$SO(6)$ sector we find that the one loop corrections to  
structure constants of gauge invariant operators is 
determined by the one loop anomalous dimension Hamiltonian in this
sector. We then evaluate the one loop corrections to structure constants for 
scalars with arbitrary number of derivatives in a given holomorphic 
direction. We find that the corrections can be characterized by
suitable derivatives on the four point 
tree function of  a massless 
scalar with  quartic coupling. 
We show that 
individual diagrams violating conformal invariance can be combined
together to restore it using a linear inhomogeneous partial
differential equation satisfied by this function. 
}

\begin{document}
\baselineskip 4ex

\section{Introduction}

By far, the most precise realization  of  field theories being
dual to  string theories  occurs in examples of the AdS/CFT
correspondence proposed by Maldacena 
\cite{Maldacena:1997re,Witten:1998qj,Aharony:1999ti}. 
Among these examples, 
the most  studied case is the duality between ${\cal N}=4$
Yang-Mills theory in four dimensions with 
gauge group $U(N)$ and type IIB string
theory on $AdS_5\times S^5$. 
Let us briefly recall the map between the basic parameters of
the string theory and ${\cal N}=4$ Yang-Mills. 
It is convenient to set the radius of $AdS$ to one so that in such units
the string length is related to the t'Hooft coupling
of the gauge theory  by
\be{diction}
\alpha' = \frac{1}{\sqrt{\lambda}} = \frac{1}{\sqrt{g_{YM}^2
N}},\;\;\;\; \;\;\;\;\;\;\; G_N = \frac{1}{N^2},
\ee
here $g_{YM}$ is
the Yang-Mills coupling constant, $\alpha'$ refers to the
string length and $G_N$ is the Newton's constant in these units
which is the effective string loop counting parameter.

The regime in which this duality has been
mostly explored is when the type IIB string theory can be approximated
by type IIB supergravity. 
To decouple all the string modes, 
the t'Hooft coupling has to be large. 
Furthermore, 
to suppress string loops we need to work at large
$N$. One can then set up a precise correspondence of gauge invariant
operators and supergravity fields. 
Another interesting limit, which has received a lot of attention
recently,
is when the t'Hooft coupling $\lambda$ is 
small but with $N$ still being large. In this limit especially 
when $\lambda$ is strictly zero, all string modes are equally important
but string loops are suppressed.
From \eq{diction} we see that $\lambda$ 
being zero implies the string length is infinity, 
the $AdS_5\times S^5$  string  sigma model is strongly coupled.
At present there are no known methods to extract any information
regarding the spectrum or the correlation functions from the strongly 
coupled sigma model. On the other hand, the dual field theory is best
understood in this limit since at $\lambda =0$ the theory is free and 
planar perturbation theory in  the 
t'Hooft coupling is sufficiently easy to
perform. This has led to many efforts in trying to 
rewrite  the spectrum of the ${\cal N}=4$ Yang-Mills theory
as a spectrum in a string theory 
\cite{Bianchi:2003wx,Beisert:2003te,Bianchi:2004xi}.
There has also been an effort  at
reconstructing the 
string theory  world sheet by rewriting the 
correlation function of gauge invariant operators of the free theory
as amplitudes in $AdS$ \cite{Gopakumar:2003ns,Gopakumar:2004qb}.

In this paper, with the motivation to find
features of the string theory 
at weak coupling Yang-Mills 
we study structure constants of certain class of gauge
invariant operators in planar ${\cal N}=4$ super Yang-Mills, at 
one loop in t'Hooft coupling. 
To indicate which features of the string theory one would
expect to see by studying the structure constants, 
we  first need to provide the  picture of the string theory 
at  $\lambda =0$ limit that we have in mind. 
From \eq{diction} we see that at $\lambda=0$
the string essentially becomes tensionless, therefore there is 
no coupling between neighboring points 
on the string which  breaks up into non interacting bits. 
In fact this picture of the string has already been
noticed in the plane wave limit \cite{Berenstein:2002jq}
and has been discussed in the context of
string theory in small radius AdS \cite{Dhar:2003fi}.
From studies of correlation functions of gauge invariant operators 
in the plane wave limit,
it is seen  that 
each Yang-Mills letter can be thought of
as a bit in a light cone gauge fixed string theory, 
and a single trace gauge invariant operator is a sequence of bits 
with cyclic symmetry
\cite{Constable:2002hw,Kristjansen:2002bb,
Verlinde:2002ig,Vaman:2002ka,Zhou:2002mi}
A universal feature of any  string field theory is that
interactions are described by  delta function overlap
of  strings. 
Therefore the structure constants of gauge invariant operators,
which in the planar limit are proportional to $1/N$, should be
seen as joining or splitting of strings.
Indeed, it is possible to formulate a bit string theory in which all
features of the two point functions  and 
structure constants of gauge invariant operators, 
including position dependence, can
be reproduced by the delta function overlap \cite{adgn:2005}.

Now let us ask the question of what would be the modifications in the
above picture when one makes $\lambda$ finite. 
From \eq{diction} we see that  rendering 
$\alpha'$ finite  would introduce 
interactions between the bits. 
At first order in $\lambda$
and in the planar limit, only nearest neighbor bits would
interact. Therefore, turning on $\lambda$ modifies  
the free propagation of
the bits in the bit string theory.
The one loop corrected 
two point function and the structure constants  
should still be determined by the geometric delta function overlap,
but with the modification in the propagation of the bits 
taken into account. 
Thus identifying the precise operator which is responsible for the
propagation of the bits at
first order in $\lambda$, should be 
sufficient to 
determine the modified two point  functions and the structure
constants at one loop.
It is this feature of Yang-Mills theory 
we hope to uncover by studying 
the structure constants.

Apart from the above motivations, from a purely field theoretic point
of view
a conformal field theory is completely specified by the
the two point functions and the structure constants of the operators.
A lot of effort have been made to understand the structure of the 
two point functions of gauge 
invariant operators  of ${\cal N}=4$ Yang-Mills in the planar limit.
In fact  the anomalous dimension  Hamiltonian at one loop in
$\lambda$ 
is known to be integrable \cite{Minahan:2002ve,Beisert:2003tq,
Beisert:2003jj,Beisert:2003yb}, 
and   signatures of integrability
in the form of the existence of an infinite number of nonlocal
conserved charges has been shown for
the world sheet theory on $AdS_5\times S^5$
\cite{Mandal:2002fs,Bena:2003wd,Vallilo:2003nx,Alday:2003zb,
Wolf:2004hp}.
Furthermore, the relation between these approaches to integrabilty
have been studied in \cite{Dolan:2003uh,Dolan:2004ps,Dolan:2004ys}.
On the other hand structure constants of operators in  ${\cal N}=4$ 
theory are considerably less explored 
\cite{Bianchi:2001cm,Okuyama:2004bd,Roiban:2004va}.
One difficulty in studying corrections to structure constants is
that one needs to find the right renormalization group
invariant quantity which characterizes the corrections 

In this paper we derive a 
simple formula which characterizes the
renormalization group invariant quantity  which determines the
corrections to structure constants of  primary gauge invariant
operators. Then we use this to study the one loop corrections to
structure constants in the scalar $SO(6)$ sector and a sector of
operators with derivatives in a given holomorphic direction. 
We find that in the $SO(6)$ sector the renormalization invariant
quantity, which determines 
the one loop correction to the structure
constants, is the one loop anomalous dimension Hamiltonian itself. 
Evaluation of the structure constants 
for operators with derivatives is
considerably more involved. 
Feynman graphs contributing to the corrections can be obtained by
a suitable combination of 
derivatives acting on the  function $\phi(r,s)$, 
which refers to  the tree level
four point function of  a massless scalar with
a quartic coupling and  $r,s$ are the
two conformal cross ratios. 
There are individual 
Feynman diagrams contributing
to the one loop corrections to structure constants 
which seem at first to violate conformal
invariance, but we find that the violating 
diagrams can be combined together using the fact that
 $\phi(r,s)$ satisfies a linear
inhomogeneous partial differential equation ensuring conformal
invariance \footnote{After completion of this work 
it was pointed out to us by G. Arutyunov, 
that similar differential equations have been studied in 
\cite{Eden:2000bk,Arutyunov:2002fh}}.

This paper is organized as follows. In section 2. we derive the
renormalization group invariant formula characterizing the
corrections to structure constants of primary operators. 
In section 3. we apply this to the 
scalar $SO(6)$ sector and show that corrections are captured by the
one loop anomalous dimension Hamiltonian. 
The fact that the anomalous dimension 
Hamiltonian captures the correction to the structure constants
was observed in \cite{Okuyama:2004bd}. 
Their observation relied on certain
examples and the statement that only the $F$ terms occur in the
Feynman diagrams. The proof given here is direct and 
the method is suitable for extension 
for  classes of operators in other sectors.
In section 4. we compute the corrections to structure
constants for operators with derivatives
in one holomorphic direction. We show that conformal invariance
in the  three point function is ensured by the differential equation
satisfied by $\phi(r,s)$. The summary of the results which enables one
to calculate the structure constants to any operator in this sector is
given in  section 4.4.
Appendix A. contains the  notations adopted in the 
paper,  Appendix B discusses the properties of the
function $\phi(r,s)$, in particular it contains the proof of the 
differential equation it satisfies. 
Appendix C. contains tables which are required
in the evaluation of the structure constants in the derivative sector.

\section{General form of structure constants at one loop}

Our aim 
in this section is to  
derive a formula which gives a renormalization
group invariant characterization of one loop corrections to structure
constants at large $N$. 
Consider a set of conformal primary operators  labelled by
$O_{i}^{\mu_1 \ldots \mu_{n_i}}$, 
here $\mu_1 \ldots \mu_{n_i}$ indicate the tensor structure of the primary
\footnote{In this paper will restrict our attention to primaries which
are tensors, but our methods can be generalized to other classes
of operators.}.
For simplicity, let us suppose the basis of  operators is such that
their one loop anomalous dimension matrix is diagonal, 
we will relax this assumption later. Then, by
conformal invariance, the general form for the 
two point function of
these operators at one loop is given by: 
\be{g2pt}
\langle O_{i}^{\mu_1 \ldots \mu_{n_i}} (x_1) 
O_{j}^{\nu_1 \ldots \nu_{n_j} } (x_2) \rangle
= \frac{ J^{\mu_1\ldots \mu_{n_i} ;
\nu_1\ldots \nu_{n_i} } }{(x_1 -x_2)^{2\Delta_i  } }
\left( \delta_{ij} + \lambda g_{ij} - \lambda \gamma_{i} \delta_{ij} 
\ln ((x_1 -x_2)^2 \Lambda^2 ) \right).
\ee
Here $J^{\mu_1 \ldots \mu_{n_i} ; \nu_1 \ldots \nu_{n_i} }$ is the 
invariant tensor 
constrained by conformal invariance and 
constructed by  products of the following
tensor:
\be{defj}
J^{\mu\nu} = \delta_{\mu\nu} - 2 \frac{ (x_1- x_2)^\mu ( x_1- x_2)^\nu
} { ( x_1 - x_2) ^2 }.
\ee
Since we are interested in the one loop correction in the planar limit, 
the expansion parameter in \eq{g2pt} $\lambda = g_{YM}^2 N /32 \pi^2$ 
is the t' Hooft coupling. In \eq{g2pt}  we have used the fact that 
it is possible to choose a basis of operators such that 
they are orthonormalized at tree level and that their anomalous
dimension matrix is diagonal. 
$\Delta_i$ are the bare dimensions and 
$\gamma_i$ refer to the anomalous dimensions of the respective
operators. For non zero tree level two point function in \eq{g2pt}
$\Delta_i =\Delta_j$ and $n_i =n_j$. 
The constant mixing matrix at one loop $g_{ij}$ is
renormalization group scheme dependent, for instance if the cut off
$\Lambda$ is scaled to $ e^\alpha \Lambda$,  the mixing matrix 
changes as follows: 
\be{cgmix}
g_{ij} \rightarrow g_{ij} - 2\alpha \gamma_i\delta_{ij}. 
\ee
The three point function of three tensor primaries is given by:
\bea{3ptdef}
\langle 
& &O_i^{\mu_1 \ldots \mu_{n_i} } (x_1) 
O_j^{\nu_1 \ldots \nu_{n_j} } (x_2)
O_k^{\rho_1 \ldots \rho_{n_k} } (x_3) \rangle 
\\ \nonumber
&=& \frac{ J^{ \mu_1 \ldots \mu_{n_i} ;\nu_1 \ldots \nu_{n_j};
\rho_1 \ldots \rho_{n_k}}  }{
|x_{12}|^{\Delta_i + \Delta_j - \Delta_k }
|x_{13}|^{\Delta_i + \Delta_k - \Delta_j }
|x_{23}|^{\Delta_j + \Delta_k - \Delta_i }
} \times \\ \nonumber
& &\left( C_{ijk}^{(0)} \left[ 1 - 
\lambda \gamma_i 
\ln | \frac{ x_{12} x_{13} \Lambda }{ x_{23}} |
- \lambda \gamma_j 
\ln | \frac{ x_{12} x_{23} \Lambda }{ x_{13}} |
- \lambda \gamma_k 
\ln | \frac{ x_{13} x_{23} \Lambda }{ x_{12}} | \right]
+ \lambda \tilde C_{ijk}^{(1)} \right),
\eea
where $x_{12} = x_1 -x_2, x_{13} = x_1 - x_3, x_{23} = x_2 - x_3$.
Note, that from large $N$ counting it is easy to see that both 
$C_{ijk}^{(0)}$ and the one loop correction $ \tilde C_{ijk}^{(1)} $
are order $1/N$. Again the constant one loop correction to the 
$\tilde C_{ijk}^{(1)}$ is 
renormalization scheme dependent, scaling $\Lambda$ by $e^\alpha
\Lambda$,  we see that: 
\be{cg3pt}
\tilde C_{ijk}^{(1)} \rightarrow \tilde C_{ijk}^{(1)} - \alpha\left(
\gamma_i C_{ijk}^{(0)} +\gamma_j C_{ijk}^{(0)} +\gamma_k
C_{ijk}^{(0)}\right).
\ee
Here there is no summation of repeated indices.
Therefore from \eq{cgmix} and \eq{cg3pt} we see that the following
combination is renormalization scheme independent
\be{cgind}
C_{ijk}^{(1)} = \tilde C_{ijk}^{(1)} 
- \frac{1}{2} g_{ii'} C_{i'jk}^{(0)} - \frac{1}{2} g_{jj'}
C_{ij'k}^{(0)} 
- \frac{1}{2} g_{kk'} C_{ijk'}^{(0)}, 
\ee
where summation over the primed indices is implied.
Essentially, the renormalization scheme independent 
one loop correction to the
structure constant is obtained by first normalizing all the two point
function to order $\lambda$.
We now write the  equation \eq{cgind} 
using an arbitrary basis of primaries.
Let the transformation matrix 
which takes the orthonormalized basis of
primaries to an arbitrary basis, be given by $U_{\alpha i}$, where
$\alpha, \beta \ldots$ label the arbitrary basis, of primaries. This
transformation is $\lambda$ independent  since it is possible to
choose a basis of operators which are orthonormalized at tree level
and their one loop anomalous dimension matrix is diagonal. The
transformation matrix $U_{\alpha i}$ satisfies the following
relations:
\be{propu}
\sum_{i} U_{\alpha i} U_{\beta i} = h_{\alpha\beta}, \;\;\;\;\;\;\;
\sum_{i} U_{\alpha i} \gamma_i U_{\beta i} = \gamma_{\alpha\beta}.
\ee
Here $h_{\alpha\beta}$ is the tree level mixing matrix and 
$\gamma_{\alpha\beta}$ is the anomalous dimension matrix at one loop.
It is usually convenient to chose a basis with $h_{\alpha\beta} =
\delta_{\alpha\beta}$, in standard literature the anomalous dimension
matrix is specified in such a basis.
But here we will work with an arbitrary basis,
performing change of  basis in \eq{cgind} we obtain:
\be{3ptbind}
C^{(1)}_{\alpha\beta\gamma} = \tilde 
C_{\alpha\beta\gamma}^{(1)}
- \frac{1}{2} g_{\alpha\alpha'} C^{(0)\alpha'}_{\;\;\beta\gamma}
-\frac{1}{2} g_{\beta\beta'} C^{(0)\;\;\beta'}_{\alpha\;\;\;\;\gamma}
- \frac{1}{2} g_{\gamma\gamma'} C^{(0)\;\;\;\;\gamma'}_{\alpha\beta},
\ee
where:
\bea{def3ptnew}
\tilde C^{(1)}_{\alpha\beta\gamma} = U_{\alpha i} U_{\beta j}
U_{\gamma k} \tilde C^{(1)}_{ijk},  &\;&\;\;
 C^{(0)}_{\alpha\beta\gamma} = U_{\alpha i} U_{\beta j}
U_{\gamma k} \tilde C^{(0)}_{ijk}, \\ \nonumber
C^{(0) \alpha}_{\;\;\beta\gamma} = h^{ \alpha\alpha'}
C^{(0)}_{\alpha'\beta\gamma}, &\;&\;\;
C^{(0)\;\; \beta}_{\alpha\;\;\;\;\gamma}
= h^{ \beta\beta'} C^{(0)}_{\alpha\beta'\gamma},
\;\;\;
C^{(0) \;\;\;\gamma}_{\alpha\beta}
= h^{\gamma\gamma'} C^{(0)}_{\alpha\beta\gamma'}, \\ \nonumber
h^{ \alpha\alpha'} h_{\alpha' \beta} &=& \delta^\alpha_\beta.
\eea
We will call the subtractions in \eq{3ptbind} as metric subtractions.

\subsection{The slicing argument}

We work towards a useful characterization of the formula given in
\eq{3ptbind}. Local
gauge invariant  operators can be constructed by  products of the 
fundamental letters of ${\cal N}=4$ Yang Mills and finally taking a
trace.  We represent a general Yang Mills letter by $W^A$, then a
gauge invariant operator is $\rm Tr( W^A W^B \cdots W^Z)$. 
The tree level contractions which contribute to 
$C^{(0)}_{\alpha\beta\gamma}$ 
of three gauge invariant primaries at the planar level 
are all possible Wick contractions which can be drawn on a plane
using the double line notation. We can represent a given contraction
by the diagram in \fig{pwick}, the corresponding double line notation is
given adjacent to it. 
\FIGURE{
\label{pwick}
\centerline{\epsfxsize=16.truecm \epsfbox{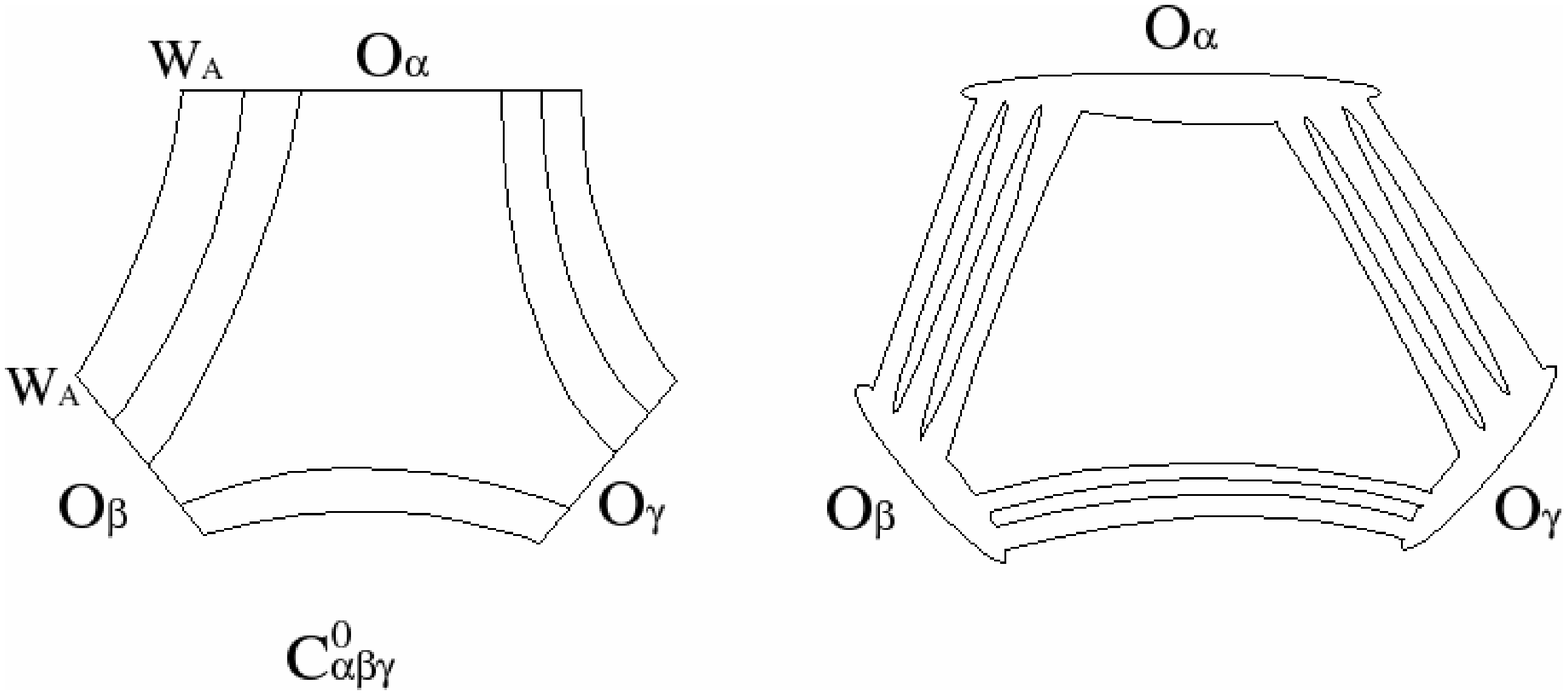}}
\caption{Planar Wick contractions contributing to
$C^{(0)}_{\alpha\beta\gamma}$}
}
In \fig{pwick} we have used single lines to represent the double
line. The lines end on letters of the operators, these are points on
the horizontal lines in the diagram.

Consider the one loop correction $\tilde C^{(1)}_{\alpha\beta\gamma}$,
contributions to this can arise from two types of terms: 
(i) two body terms represented by  $U_{\alpha\beta}, 
U_{\alpha\gamma}$ and $U_{\beta\gamma}$  in
\fig{2bdy}
(ii) genuine three body terms represented by 
$U^{\alpha}_{\beta\gamma}, 
U^{\beta}_{\gamma\alpha},
U^{\gamma}_{\alpha\beta}$  as shown in \fig{3bdy}.
\FIGURE{
\label{2bdy}
\centerline{\epsfxsize=16.truecm \epsfbox{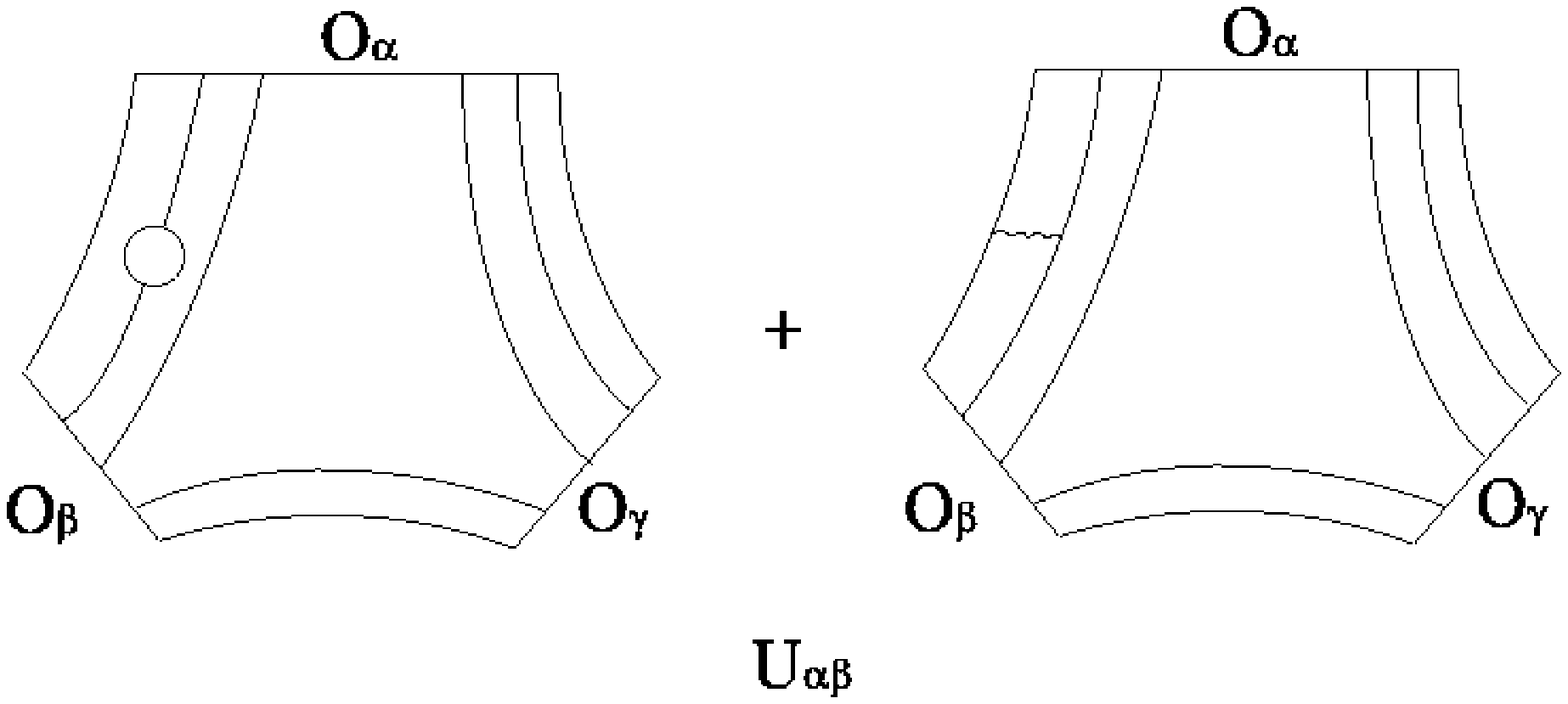}}
\caption{A generic diagram contributing to $U_{\alpha\beta}$ }
}
\FIGURE{
\label{3bdy}
\centerline{\epsfxsize=16.truecm \epsfbox{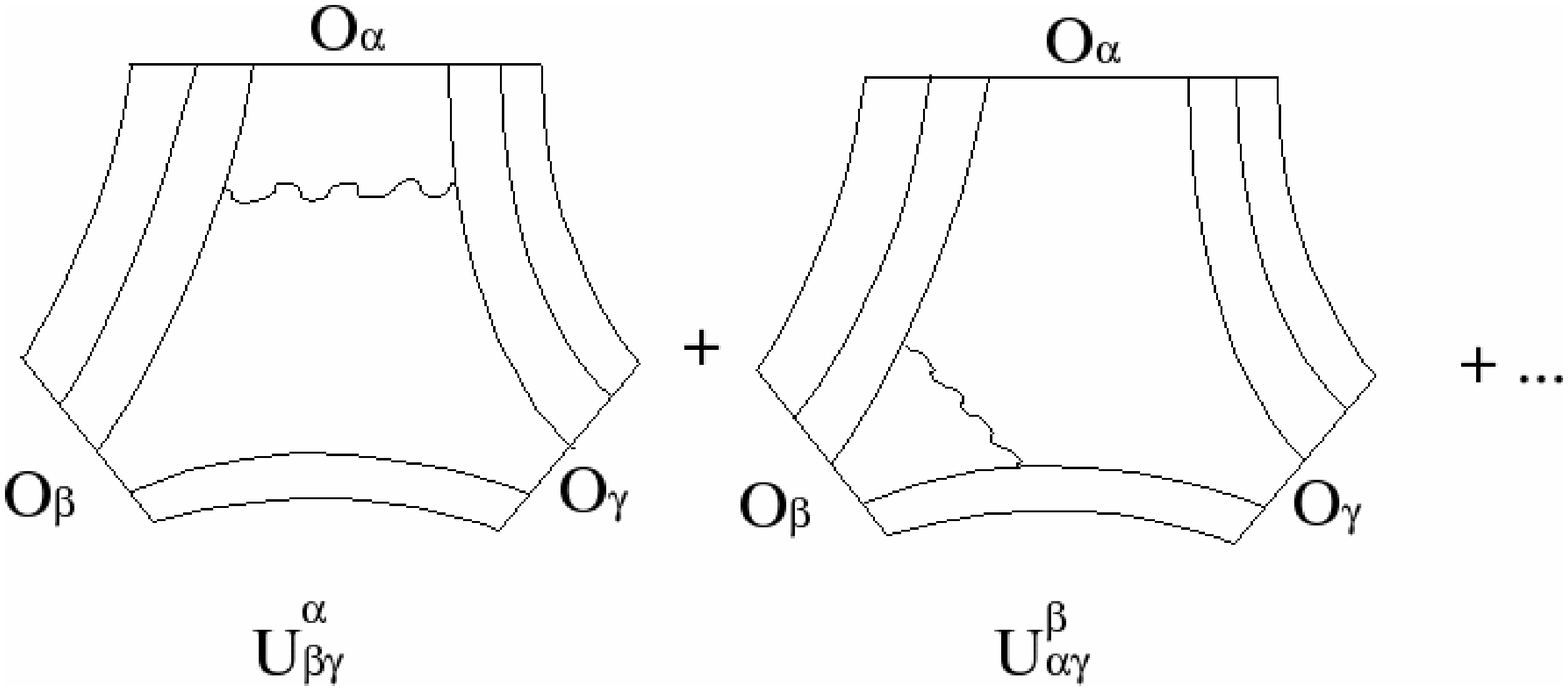}}
\caption{Diagrams contributing to $U^\alpha_{\beta\gamma}$ and
$U^\beta_{\alpha\gamma}$ }
}
As we are 
interested in planar corrections at one loop, it is easy to see that 
the two body interactions can  occur only  between nearest neighbour
letters of 
any two of the operators with the remaining contractions performed at
the free level. There is an exception to this rule, when the structure
constant of interest is length conserving, for instance when say, the length 
of  operator $O_\alpha$ equals the sum of the lengths of the 
operators $O_\beta$ and $O_\gamma$. We will discuss this case later in 
the paper, but for now 
and for most of the discussions in this paper 
we assume that the structure constants of interest 
are length non-conserving. 
Two body interactions can also consist of 
planar self energy interactions
between letters of any two different operators, 
and the rest of the operators contracted with free Wick contractions.
Thus $U_{\alpha\beta}$
represents the sum of all the constants due to all possible nearest
neigbour interactions among operators $O_\alpha$ and $O_\beta$, 
and all possible
constants from the 
self energy interactions between letters of these operators. 
A similar definition holds for $U_{\alpha\gamma}$ and $U_{\beta\gamma}$. 
The genuine three body term $U^{\alpha}_{\beta\gamma}$ consists of 
constants from all possible 
interactions between any two nearest neighbour letters of
the operator $O_\alpha$ and two letters of operators $O_\beta$ and 
$O_\gamma$ such
that all contractions are planar. An example of such an interactions
are shown in \fig{3bdy}. 
It is easy to see from this diagram that one is forced to choose nearest
neighbour letters in operator $O_\alpha$ to ensure that the interaction is 
planar. 
Similar definitions hold for $U_{\gamma\alpha}^\beta,
U^\gamma_{\alpha\beta}$. From these definitions we have:
\be{3ptbrek}
\tilde C^{(1)}_{\alpha\beta\gamma} = U^{\alpha}_{\beta\gamma}
+ U^\beta_{\gamma\alpha} + U^\gamma_{\alpha\beta} +
U_{\alpha\beta} + U_{\beta\gamma} + U_{\gamma\alpha}.
\ee

We show now that  the two body terms of
$\tilde C^{(1)}_{\alpha\beta\gamma}$ cancel with the  metric
subtractions in the equation \eq{3ptbind}.
Consider a generic two body interaction in $U_{\alpha\beta}$, imagine
slicing the diagram as in \fig{slice}  
by inserting a complete set of operators
$O_{\alpha'}$. 
\FIGURE{
\label{slice}
\centerline{\epsfxsize=16.truecm \epsfbox{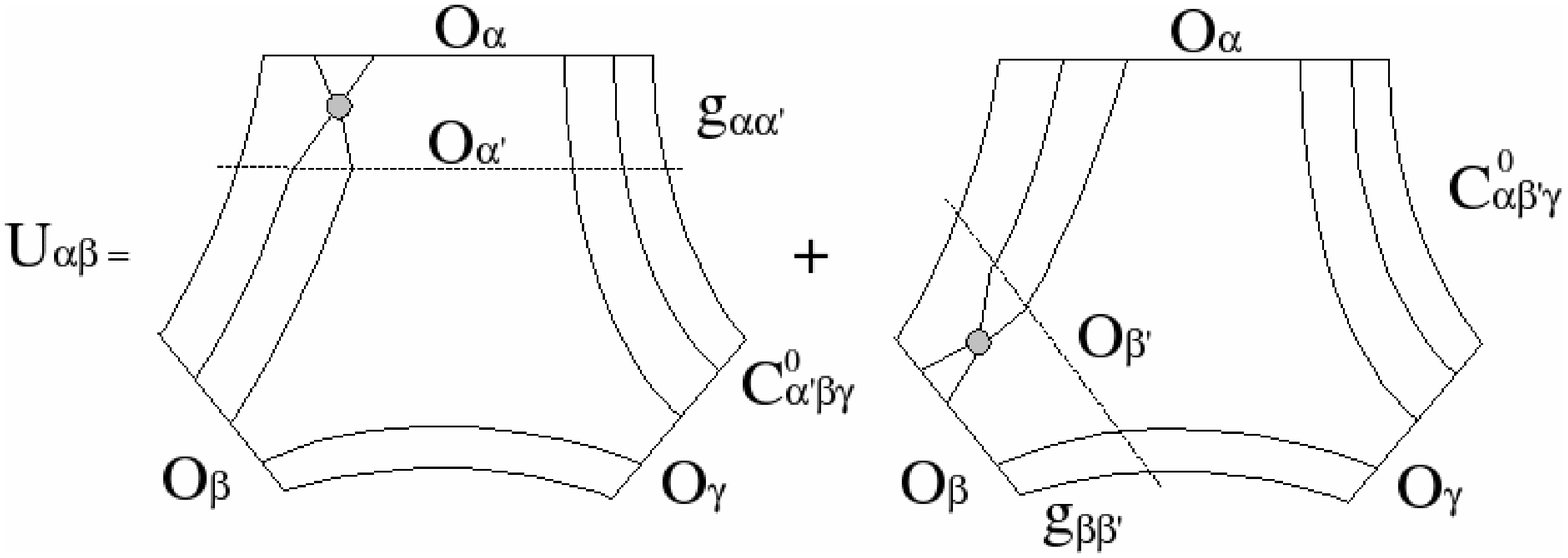}}
\caption{The slicing argument }
}
Thus the diagram decomposes into
two halves, the upper half which contains the one loop corrections 
which can now be viewed as  contributions 
to the  one loop correction
$g_{\alpha\alpha'}$. The
lower half which is just the tree level structure constant
$C^{(0)\alpha'}_{\;\; \beta\gamma}$. 
From this slicing we see that
exactly the same one loop interaction term occurs in 
$g_{\alpha\alpha'} C^{(0)\alpha'}_{\;\;\beta\gamma}$ 
\footnote{In the first
diagram  in \fig{slice} we have shown only one interaction diagram
which on slicing gives a contribution to $g_{\alpha\alpha'}$, other
contributions to $g_{\alpha\alpha'}$ also comes from interactions
in lines running between $O_\alpha$ and $O_\beta$ in this slicing.}. 
Now, slice the same diagram as indicated in the second figure of 
\fig{slice} 
by inserting a complete set of 
operators $O_{\beta'}$. The one loop correction can be seen as a term in 
$g_{\beta\beta'}$, while the rest of the diagram as the tree level
structure constant $C^{(0) \;\;\beta'}_{\alpha\;\;\gamma}$.
Thus  this diagram also
occurs in 
$g_{\beta\beta'} C^{(0)\;\;\beta'}_{\alpha\;\;\gamma}$. In 
\eq{3ptbind}, the metric subtractions 
$g_{\alpha\alpha'} C^{(0)\alpha'}_{\;\;\beta\gamma}$ 
and 
$g_{\beta\beta'} C^{(0)\;\;\beta'}_{\alpha\;\;\gamma}$ 
are weighted by a factor of $1/2$, thus 
we conclude that a generic two body interaction
in $U_{\alpha\beta}$ is canceled off by the subtractions in 
\eq{3ptbind}. This cancellation includes  both the nearest
neighbour two body interactions as well as the self energy type of
interactions which we have not shown in \fig{slice}.
Similar reasoning can be used to conclude that 
the all the constants in the two body terms $U_{\beta\gamma}$ and
$U_{\gamma\alpha}$ also are canceled by the metric subtractions in 
\eq{3ptbind}. 

From the slicing argument we see that 
the constants from a genuine three body terms in 
$U^{\alpha}_{\beta\gamma}, 
U^{\beta}_{\gamma\alpha}, U^{\gamma}_{\alpha\beta}$ 
cannot be canceled of the metric subtractions. 
Thus these terms and the 
corresponding subtraction in \eq{3ptbind} is what is left behind. 
This is indicated in the \fig{leftover}. 
\FIGURE{
\label{leftover}
\centerline{\epsfxsize=16.truecm \epsfbox{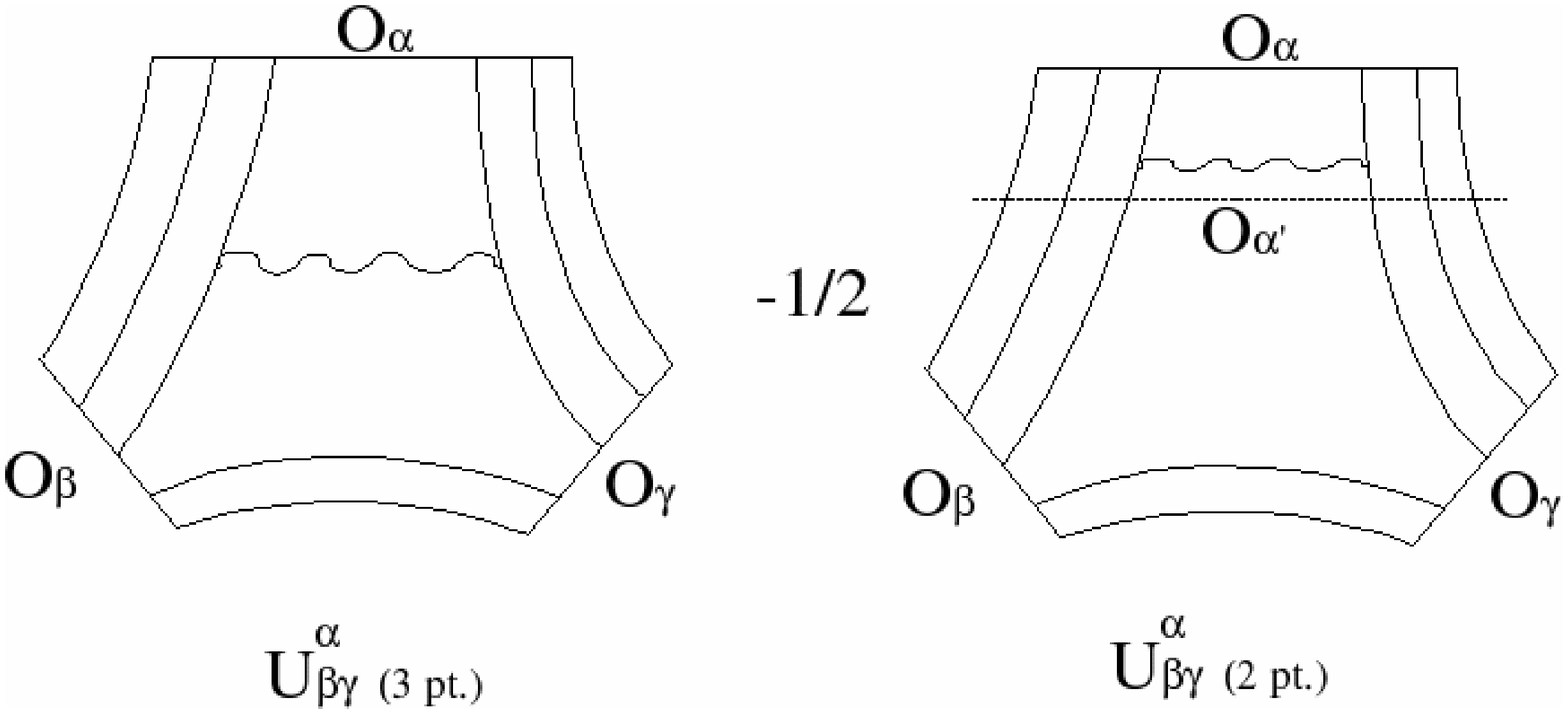}}
\caption{Renormalization scheme independent contribution}
}
Therefore computation of 
$C^{(1)}_{\alpha\beta\gamma}$ reduces to the evaluation of
constants from diagrams with 4 letters: 2 letters on one operator, 
say $O_\alpha$, and the remaining 2 letters on operators $O_\beta$ and 
$O_\gamma$. From this we subtract half the constants which occur when
the same diagram is thought of as the two body interaction, that is 
2 letters on one operator say $O_\alpha$  and the 
remaining $2$ letters on the operator $O_\alpha'$. Summing over all
such contributions gives $C^{(1)}_{\alpha\beta\gamma}$. 
We write this compactly as 
\bea{3ptfin}
C^{(1)}_{\alpha\beta\gamma} &=&
\left( U^{\alpha}_{\beta\gamma} ({\rm 3pt } ) - 
\frac{1}{2} U^{\alpha}_{\beta\gamma} ({\rm 2pt } )  \right)
+ \left( U^{\beta}_{\gamma\alpha} ({\rm 3pt} ) 
-\frac{1}{2} U^{\beta}_{\gamma\alpha} ({\rm 2pt} ) \right) \\
\nonumber
&+& \left( U^{\gamma}_{\alpha\beta} ({\rm 3pt})
-\frac{1}{2} U^{\gamma}_{\alpha\beta} ({\rm 2pt}) \right)
\eea
Here $U^{\alpha}_{\beta\gamma} (3{\rm pt})$ contains constants from
genuine three body interactions, that is there are no self energy
diagram. $U^{\alpha}_{\beta\gamma}( 2{\rm pt})$ contains the constants
from the same diagrams but now thought of as occurring in a two point
function, to emphasize again,  this also has no self energy diagrams.
Therefore, to compute one loop corrections to structure constants for any
arbitrary operator it is sufficient to give the 
one loop corrections  occurring in 
the computation of any $4$ Yang Mills letters, 
firstly thought of as genuine 3 body interaction and then thought of
as a two body interaction. 

\subsection{An example}

We illustrate the slicing argument using a simple example by explicitly
evaluating all the terms occurs in \eq{3ptbind} 
and showing that it reduces to \eq{3ptfin}.
Consider the  structure constant when the operators are given by
\be{defop}
O_\alpha = O_\beta= O_\gamma = \frac{1}{N} {\rm Tr} (Z\bar{Z}).
\ee
Here $Z$ is a
complex scalar in the one of the Cartan of  $SO(6)$, for instance
$ Z = \frac{1}{\sqrt{2}} ( \phi^1 + i \phi^2)$.  
Thus the $Z$ , $\bar{Z}$ 
Wick contraction is normalized to $1$, which implies that the 
tree level two point function $h_{\alpha\alpha}$  is normalized to $1$. 
Evaluating the tree level structure constant 
we obtain $C_{\alpha\alpha\alpha} = 2/N$. 

Now consider the one loop corrections to the structure constants. The 
two body terms consists only of self energy diagrams, these are given
by 
\be{2body}
U_{\alpha\beta}+ U_{\beta\gamma} + U_{\gamma\alpha} 
= \frac{\lambda}{N} \left(  2 S_{\alpha\beta} + 2S_{\alpha\gamma} +
2S_{\beta\gamma}\right)  =
\frac{\lambda}{N} 6 S.
\ee
The subscripts in the $S$ are just used to indicate the origin of the 
constants from the self
energy diagrams, for instance there are two self energy diagrams between
the $Z$ and $\bar Z$ of the $O_\alpha$ and $O_\beta$. 
Since all the self energy diagrams are same they can be summed to give $6S$. 
We have also
kept track of the order of the t' Hooft coupling and $N$. 
The genuine three body terms are
\be{3body}
U^\alpha_{\beta\gamma} + 
U^\beta_{\gamma\alpha} + 
U^\gamma_{\alpha\beta}  
= \frac{\lambda}{N} \left[
4 H(\alpha;\beta\gamma) + 
4 H(\beta;\gamma\alpha) + 4 H(\gamma;\alpha\beta) \right] 
= \frac{\lambda}{N} 12 H( {\rm 3pt}).
\ee
Here the $H$ basically refers to the constant from the diagram with 
$Z$ and $\bar{Z}$ on one operator and with $\bar{Z}$ and $Z$ on the
remaining two operators. The labels in each of the $H$ just refer to
which of the operator has the two letters and which of the rest has a
letter each. The factor 4 arises out of the combinatorics of the
diagrams. Therefore we have
\be{ex3pt}
\tilde C^{(1)}_{\alpha\alpha\alpha}  
=\frac{\lambda}{N} \left[  6S + 12 H( {\rm 3pt}
) \right]. 
\ee

Now we subtract out the metric contributions in \eq{3ptbind}.
We have to sum over all the metric contributions
$g_{\alpha\beta'}C^{(0) \beta'}_{\;\;\alpha\alpha}$, 
but this sum reduces to 
evaluating only one term when $\beta' =\alpha$, this is because all other 
tree level structure constants vanish. 
Now $g_{\alpha\alpha}$ is given by
\be{forg}
g_{\alpha\alpha} = \lambda[ 2 S + 2 H({\rm 2pt}) ],
\ee
thus we see that
\bea{opf}
C^{(1)}_{\alpha\alpha\alpha} &=& \tilde C^{(1)}_{\alpha\alpha\alpha}
- \frac{1}{2} 3 g_{\alpha\alpha}
C^{(0)\alpha}_{\;\;\alpha\alpha}, \\ \nonumber
&=&  12 \frac{\lambda}{N} \left( H( {\rm 3pt}) - \frac{1}{2} H({\rm
2pt}) \right),
\eea
where we have used \eq{ex3pt} , \eq{forg} and substituted the value of
$C^{(0)\alpha}_{\;\;\alpha\alpha} =
h^{\alpha\alpha}C^{(0)}_{\alpha\alpha\alpha} = 2/N$. 
Note that the self energies which are  the only two body 
terms in $\tilde C^{(1)}_{\alpha\alpha\alpha}$ 
have canceled on subtracting the 
metric contributions.
The last formula in \eq{opf} is precisely
the equation one would have obtained if  one uses the formula in
\eq{3ptfin}.

\section{The scalar $SO(6)$ sector}

Consider three  operators belonging only
to the scalar $SO(6)$ sector given by 
\bea{defsop}
O_\alpha = \frac{1}{N^{l_\alpha/2} }
{\rm Tr} ( \phi^{i_1} \phi^{i_2} \ldots \phi^{i_{l_\alpha}} )
\\ \nonumber
O_\beta = \frac{1}{N^{l_\beta/2} }
{\rm Tr} ( \phi^{j_1} \phi^{j_2} \ldots \phi^{j_{l_\beta}} )
\\ \nonumber
O_\gamma = \frac{1}{N^{l_\gamma/2} }
{\rm Tr} ( \phi^{k_1} \phi^{k_2} \ldots \phi^{k_{l_\alpha}} )
\eea
In this section we show that the renormalization scheme independent
correction to the structure constants of this
class of operators is essentially dictated by the 
anomalous dimension Hamiltonian. 
The invariant one loop correction is  given by
\be{s3pt}
C^{(1)}_{\alpha\beta\gamma} =
\sum_{a,b,c} {\cal H}^{i_a i_{a+1} }_{j_{b+1} k_c} {\cal I}  +
\sum_{a,b,c} {\cal H}^{j_b j_{b+1} }_{k_{c+1} i_a} {\cal I}  +
\sum_{a,b,c} {\cal H}^{k_c k_{c+1} }_{i_{a+1} j_b} {\cal I} 
\ee
where ${\cal H}$ is the anomalous dimension Hamiltonian given by 
\cite{Minahan:2002ve,Beisert:2003tq}
\be{anhamil}
{\cal H}^{ij}_{kl} = 2 \delta^j_k\delta^i_l - 2 \delta^i_k\delta^j_l -
\delta^{ij}\delta_{kl}.
\ee
${\cal I}$  in \eq{s3pt} refers to the remaining free planar contractions as
shown in \fig{leftover}.
The summation over $a, b, c$ runs over all distinct
cyclic permutations of the diagram over the indices $i$, $j$ and $k$ of
the three operators. In \eq{s3pt} and 
through out the rest of the paper we will suppressed the 
$\lambda/N$ factor
which occurs in the normalization of the one loop corrected structure
constant.

From the slicing argument it is clear that to show \eq{s3pt} one needs
to evaluate  the following
\bea{basic}
\nonumber
& & \left( U^{i_a i_{a+1} }_{j_{b+1}   k_c} ({\rm 3pt}) - 
\frac{1}{2} U^{i_a i_{a+1} }_{j_{b+1} k_c} ({\rm 2pt})\right)
\delta^{j_b}_{k_{c+1}}
+
\left( U^{j_b j_{b+1} }_{k_{c+1}   i_a} ({\rm 3pt}) - 
\frac{1}{2} U^{j_b j_{b+1} }_{k_{c+1} i_a} ({\rm 2pt})\right)
\delta^{k_c}_{i_{a+1}} \\ 
&+&
\left( U^{k_c k_{c+1} }_{i_{a+1}   j_b} ({\rm 3pt}) - 
\frac{1}{2} U^{k_c k_{c+1} }_{i_{a+1} j_b} ({\rm 2pt})\right)
\delta^{i_a}_{j_{b+1}}
\eea
In the above formula 
$U^{i_a i_{a+1}}_{j_{b+1} k_c}( {\rm 3pt} )$ 
refers to the
constant from the diagram with  adjacent 
letters $\phi^{i_a}$, $\phi^{i_{a+1}}$ on the operator $O_\alpha$ and
the letters $\phi^{j_{b+1}}$ and $\phi^{k_c}$ on the operators
$O_\beta$ and $O_\gamma$ respectively. While 
$U^{i_a i_{a+1}}_{j_{b+1} k_c}( {\rm 2pt} )$ refers to the constant of
the same diagram but thought of as an interaction in  a two point
calculation. 
A similar definition holds for the rest of the $U$'s in \eq{basic}.
We have written down the Kr\"{o}necker delta in each of the terms in
\eq{basic} to denote the adjacent free Wick contractions. 
The terms in \eq{basic} are the generic terms that occur when the
equation \eq{3ptfin} is applied to the $SO(6)$ scalars. 
We will show that after evaluation of the terms in \eq{basic}, the
expression reduces to that given in \eq{s3pt}, essentially the $U$'s
are replaced by the anomalous dimension Hamiltonian ${\cal H}$. 

The claim that the anomalous dimension Hamiltonian dictates the 
renormalization scheme independent corrections to the structure
constants might at first be puzzling to the reader. 
The anomalous dimension
Hamiltonian arises after including self energy diagrams 
\cite{Minahan:2002ve,Beisert:2003tq} but as we 
have emphasized 
in the previous section, the renormalization scheme independent 
corrections to the three point functions do not contain any two body
terms and  in particular, there are no self energy terms. 
Therefore there is an apparent puzzle: we show below, the fact that
even the corrections to structure constants are determined by 
the anomalous dimension Hamiltonian is due to important cancellations
which take place in the evaluation of \eq{basic}

\subsection{Evaluation of corrections to structure constants }

We first evaluate the diagram $U^{ij}_{kl}$ thought of as a 3 body
term.
Consider $4$ scalars, 2 of them with indices $i$ and $j$ being nearest
neighbour letters on the operator $O_\alpha$, 
As they belong to the same operator they are at the same position.
But to regularize the resulting diagrams we 
use the method of point split regularization, therefore we split them 
such that the operator with index $i$ is at $x_1$, while the operator
with index $j$ is at $x_2$ with 
$x_2 - x_1 = \epsilon$, and $\epsilon \rightarrow 0$. 
Let
the index $k$ label the  letter of operator $O_\beta$ at position $x_3$ and
the index $l$ label the letter of operator $O_\gamma$ at position 
$x_4$.

The two process that contribute to $U^{ij}_{kl}({\rm 3 pt})$ are 
the quartic interaction of scalars and the interaction due to the
intermediate gauge exchange. Therefore
\be{defu}
U^{ij}_{kl} = Q^{ij}_{kl} + G^{ij}_{kl},
\ee
where $Q^{ij}_{kl}$ refers to the quartic interaction and
$G^{ij}_{kl}$ refers to the gauge exchange diagram. 
Evaluating each of the diagrams we obtain:
\be{squart}
Q^{ij}_{kl} = 
\lim_{x_2 \rightarrow x_1} 
\left( 2 \delta^{j}_k \delta^i_l -
\delta^i_k \delta^j_l - \delta^{ij}\delta_{kl} \right)
\frac{1}{x_{13}^2 x_{24}^2 } \phi(r, s),
\ee
here the $SO(6)$ structure arises from the quartic potential of 
the scalars in ${\cal N}=4$ super Yang-Mills, 
$\phi(r,s)$ is the quartic tree interaction 
given by 
\be{defiphi}
\int d^4u \frac{1}{(x_1 -u)^2 (x_2-u)^2 (x_3 -u)^2 (x_4 -u)^2 } = 
 \frac{\pi^2 \phi(r,s)} {x_{13}^2 x_{24}^2 },
\ee
and 
$r$ and $s$ are the conformal cross ratios given by
\be{defrs}
r= \frac{x_{12}^2 x_{34}^2 }{x_{13}^2 x_{24}^2}, \;\;\;\;
s= \frac{x_{14}^2 x_{23}^2 }{x_{13}^2 x_{24}^2}. \;\;\;\;
\ee
Note that as $x_2\rightarrow x_1$, $r\rightarrow 0$ and
$s\rightarrow1$. Therefore to evaluate the limit in \eq{squart} we can
use the expansion 
of $\phi(r,s)$ given in \eq{usexp}, substituting this expansion in
\eq{squart} we obtain
\be{fsquart}
Q^{ij}_{kl} =
\left( 2 \delta^{j}_k \delta^i_l -
\delta^i_k \delta^j_l - \delta^{ij}\delta_{kl} \right) 
\frac{1}{x_{13}^2 x_{14}^2 } \left( \ln (\frac{x_{13}^2 x_{14}^2
}{x_{34}^2 \epsilon^2 } ) +2 \right),
\ee
where we have also kept the log term for completeness.
The gauge interaction is given by 
\be{gint}
G^{ij}_{kl} = \lim_{x_2\rightarrow x_1}  \delta^{i}_k\delta^j_l
H
\ee
where
\be{defh}
H = 
(\del_1-\del_3) \cdot
(\del_2-\del_4) \int \frac{ d^4 u d^4v}{\pi^2 (2\pi)^2} \frac{1}{ 
(x_1 -u)^2 (x_3 -u)^2} \frac{1}{ (u -v)^2} 
\frac{1}{ (x_2 -v)^2 (x_3-v)^2 }.
\ee
It can be shown that $H(x_1,x_2,x_3,x_4)$ 
in the above expression can be rewritten 
entirely in terms of $\phi(r,s)$ by the following identity used in  
\cite{Beisert:2002bb}:
\bea{heq}
H  &=& E + C_1 + C_2 + C_3 + C_4, \\ \nonumber
&=&
(r-s) \frac{1}{x_{13}^2 x_{24}^2 } \phi(r,s)  \\ \nonumber
&+& (s'-r') \frac{\phi(r',s')}{x_{13}^2 x_{24}^2 }   \;\;{\rm with}
\;\;
r' = \frac{x_{34}^2}{x_{24}^2}, s' = \frac{x_{23}^2}{x_{24}^2}
;\;1\rightarrow \infty \;\;{\rm collapse} 
\\ \nonumber
&+& (s'-r') \frac{\phi(r',s')}{x_{13}^2 x_{24}^2 }   \;\;{\rm with}
\;\;
r' = \frac{x_{34}^2}{x_{13}^2}, s' = \frac{x_{14}^2}{x_{13}^2}
;\;2\rightarrow \infty \;\;{\rm collapse} 
\\ \nonumber
&+& (s'-r') \frac{\phi(r',s')}{x_{13}^2 x_{24}^2 }   \;\;{\rm with}
\;\;
r' = \frac{x_{12}^2}{x_{24}^2}, s' = \frac{x_{14}^2}{x_{24}^2}
;\;3\rightarrow \infty \;\;{\rm collapse} 
\\ \nonumber
&+& (s'-r') \frac{\phi(r',s')}{x_{13}^2 x_{24}^2 }   \;\;{\rm with}
\;\;
r' = \frac{x_{12}^2}{x_{13}^2}, s' = \frac{x_{23}^2}{x_{13}^2}
;\;4\rightarrow \infty \;\;{\rm collapse}. 
\eea
$E, C_1, C_2, C_3, C_4$ are defined respectively by the remaining
lines of the above equation. 
We have labelled $r'$ and $s'$ that occur in the second line of the above
equation by $1\rightarrow \infty$ collapse since these values are 
obtained by taking the indicated limit in $r$ and $s$ 
given in \eq{defrs}. All other values of $r'$ and $s'$ are obtained
using the corresponding limits mentioned above. We will refer to these
terms as collapsed diagrams. 
On substituting  \eq{heq} in the formula for the gauge interaction
given in \eq{gint} we need to take the limit $x_2\rightarrow x_1$. 
Under this limit $r' \rightarrow 0, s'\rightarrow 1$ for the 
$C_3$ and $C_4$ collapsed diagrams, but the 
$r'$ and $s'$ of the remaining
$C_1$ and $C_2$  collapses do not
tend of these values. On examining the expansion of $\phi(r', s')$ given
in \eq{usexp} we see that these collapsed diagrams do not reduce to 
either logarithms or constants under the limit
$x_2\rightarrow x_1$, but remain  nontrivial functions. 
Thus the collapses $C_1$ and $C_2$  seem
to violate conformal invariance, since conformal invariance of the 
3 point function predicts that the one loop correction terms must be
either logarithms or constants. We will call these collapses dangerous
collapses. 
However in the next subsection 
we will show  that on summing over all the terms given in
\eq{basic}, these dangerous collapses cancel leaving behind only
logarithms or constants.  For the present, let us assume that these
collapses cancel and evaluate the remaining terms, they are given by
\bea{gintan}
G^{ij}_{kl} ({\rm 3pt}) &=&  \delta^i_k\delta^j_l 
\left( 
 - \frac{1}{x_{13}^2 x_{14}^2} \left[ \ln \left(\frac {x_{13}^2 x_{14}^2
}{x_{34}^2 \epsilon^2} \right) + 2  \right] \right.  \\ \nonumber
&+&
\left.  \frac{1}{x_{13}^2 x_{14}^2 } \left[\ln\left( 
\frac{x_{14}^2}{\epsilon^2} \right) + 2 \right]
+  \frac{1}{x_{13}^2 x_{14}^2 }\left[ \ln\left( 
\frac{x_{13}^2}{\epsilon^2} \right) + 2 \right] \right).
\eea
The first term in the square bracket is obtained by taking the limit
$x_2\rightarrow x_1$ in the first term $E$ of  \eq{heq} and the last two
terms are obtained by taking the same limit in the
$C_3$ and $C_4$ collapsed diagrams of
\eq{heq}. Here we have ignored the 
$C_1$ and $C_2$ collapses of of \eq{heq}, as we will show 
that in the combination in \eq{basic} they cancel.
Combining all the constants to write $U^{ij}_{kl}( {\rm
3pt})$ we obtain
\be{u3ptfin}
U^{ij}_{kl}({\rm 3pt} ) =   \left[ 
2 \left( 2 \delta^{j}_k \delta^i_l -
\delta^i_k \delta^j_l - \delta^{ij}\delta_{kl} \right)  +
(-2 +2 +2) \delta^i_k \delta^j_l \right].
\ee
In the second term we have written the constant  contributions from the
first term in \eq{gintan} and the two collapses separately. 

We now evaluate $U^{ij}_{kl}({\rm 2pt})$: the calculation
is similar to the $3$ body case, except that we also need to take the
limit $x_4 - x_3 = \epsilon$ and $\epsilon\rightarrow 0$. This is
because  in the 
present calculation the letters $\phi^k$ and $\phi^l$ are
nearest neighbours on the same operator. Going through the same steps we
obtain the following contributions for the quartic term
\be{q2pt}
Q^{ij}_{kl} ({\rm 2pt}) = 
\lambda 
2 \left( 2 \delta^{j}_k \delta^i_l -
\delta^i_k \delta^j_l - \delta^{ij}\delta_{kl} \right) . 
\ee
This contribution is identical to the case of the 3 body calculation.
For the gauge exchange interaction, all the $4$ collapses, 
including $C_1$ and $C_2$,  will give
rise to logarithms and constants. 
This is because under  the 
limit $x_4\rightarrow x_3$, the corresponding $r'$ and $s'$ of 
$C_1$ and $C_2$ tends to $0$ and $1$ respectively.
Therefore the constants from the
collapses will be twice that of the 3 body calculation.
This is is given by
\be{g2ptc}
G^{ij}_{kl} ({\rm 2pt}) = 
 (-2 +2 +2+2+2) \delta^i_k \delta^j_l ,
\ee
where we have separated out the contribution of $E$ in
\eq{heq} and the 4 collapses. Thus the sum of quartic interaction and
the gauge exchange to the two body terms is given by 
\be{ufin2pt}
U^{ij}_{kl} ({\rm 2pt}) = 
2 
\left( 2 \delta^{j}_k \delta^i_l -
\delta^i_k \delta^j_l - \delta^{ij}\delta_{kl} \right)  +
 (-2 +2 +2+2+2) \delta^i_k \delta^j_l .
\ee
With all the ingredients in place, we can evaluate the renormalization
scheme independent correction to the structure constant.
This is given by
\bea{finind}
U^{ij}_{kl}({\rm 3pt} ) -\frac{1}{2} U^{ij}_{kl} ({\rm 2pt}) 
&=&  \left( 2 \delta^{j}_k \delta^i_l -2
\delta^i_k \delta^j_l - \delta^{ij}\delta_{kl} \right),   \\ \nonumber
&=&  {\cal H}^{ij}_{kl},
\eea
where we have substituted \eq{u3ptfin} and \eq{ufin2pt}. Note that
since the constant contribution of the collapses in the 2 body 
diagram are double that of the 3 body, they cancel in the renormalization
scheme independent combination.  
The gauge exchange diagram finally just contributes an additional
$-\delta^{i}_k\delta^j_l$ to give precisely 
the anomalous dimension Hamiltonian.
Substituting \eq{finind} in \eq{basic} and summing over all possible
planar contractions we will obtain \eq{s3pt} which is what we set out
to prove.  

Let us compare this calculation with the
anomalous dimension calculation of \cite{Minahan:2002ve} and
\cite{Beisert:2003tq}. There 
one focuses on the  terms proportional to the logarithm 
of  the quartic, the gauge exchange and
the self energy diagrams.
The way the Hamiltonian ${\cal H}$
appears is because the self energy contributions cancel 
all the 4 collapsed diagrams of the gauge exchange leaving
behind only the quartic $Q$ and the diagram $E$, which results in 
the anomalous dimension Hamiltonian ${\cal H}$. 
As we have seen the appearance of the anomalous dimension calculation 
in the one loop calculation of the structure constants is 
entirely due to a different mechanism.

\subsection{Cancellation of the dangerous collapsed diagrams}

In this subsection we show that the dangerous collapses in \eq{heq}
cancel out when one adds all the three terms in \eq{basic}.
The dangerous collapses when two of the indices $i_a$ and $i_{a+1}$ 
are on the same operator $O_\alpha$ is given by
\bea{dc1}
D(1;34) &=& \lim_{x_2\rightarrow x_1} 
\delta^{i_a}_{j_{a+1}} \delta^{i_{a+1} }_{k_a}
\delta^{j_a}_{k_{a+1}} \times \\ \nonumber
& & \left(
 (s'-r') \frac{\phi(r',s')}{x_{13}^2 x_{24}^2 }   \;\;{\rm with}
\;\;
r' = \frac{x_{34}^2}{x_{24}^2}, s' = \frac{x_{23}^2}{x_{24}^2}
;\;1\rightarrow \infty \;\;{\rm collapse}  \right.
\\ \nonumber
 &+& \left. (s'-r') \frac{\phi(r',s')}{x_{13}^2 x_{24}^2 }   \;\;{\rm with}
\;\;
r' = \frac{x_{34}^2}{x_{13}^2}, s' = \frac{x_{14}^2}{x_{13}^2}
;\;2\rightarrow \infty \;\;{\rm collapse} \right).
\eea
The dangerous collapse when the indices $j_a$ and $j_{a+1}$ are on the
same operator $O_\beta$ is given by
\bea{dc2}
D(3;41) &=& \lim_{x_2 \rightarrow x_3}
\delta^{i_a}_{j_{a+1}} \delta^{i_{a+1} }_{k_a}
\delta^{j_a}_{k_{a+1}} \times \\ \nonumber
& & \left(
 (s'-r') \frac{\phi(r',s')}{x_{13}^2 x_{24}^2 }   \;\;{\rm with}
\;\;
r' = \frac{x_{14}^2}{x_{34}^2}, s' = \frac{x_{13}^2}{x_{34}^2}
;\;2\rightarrow \infty \;\;{\rm collapse}  \right.
\\ \nonumber
 &+& \left. (s'-r') \frac{\phi(r',s')}{x_{13}^2 x_{24}^2 }   \;\;{\rm with}
\;\;
r' = \frac{x_{14}^2}{x_{12}^2}, s' = \frac{x_{24}^2}{x_{12}^2}
;\;3\rightarrow \infty \;\;{\rm collapse} \right).
\eea
Note that, here the limit is such $x_2\rightarrow x_3$, this
is because two letters are on operator $O_\beta$
which is at $x_3$. The index structure is identical to that of
previous case in \eq{dc1}. Finally, the values of $r'$ and $s'$ is
such that the on taking the limit in \eq{dc2} and \eq{dc1}, the last
line of the \eq{dc2} identically cancels the 1st line of \eq{dc1} when
one uses the fact $\phi(r,s)$ is a symmetric function in $r$ and $s$
\footnote{$\phi(r,s) = \phi(s,r)$ is shown in appendix B.}.
Basically the $r'$ and $s'$ of the  collapse
$2\rightarrow\infty$ of \eq{dc1} exchanges with that of the dangerous
collapse $3\rightarrow\infty$  of \eq{dc2}.
Let us now write the dangerous collapses when the indices $k_a$ and
$k_{a+1}$ are on operator $O_\gamma$ which is at position $x_4$.
\bea{dc3}
D(4;13) &=& \lim_{x_2\rightarrow x_4}
\delta^{i_a}_{j_{a+1}} \delta^{i_{a+1} }_{k_a}
\delta^{j_a}_{k_{a+1}} \times \\ \nonumber
& & \left(
 (s'-r') \frac{\phi(r',s')}{x_{13}^2 x_{24}^2 }   \;\;{\rm with}
\;\;
r' = \frac{x_{13}^2}{x_{34}^2}, s' = \frac{x_{14}^2}{x_{34}^2}
;\;2\rightarrow \infty \;\;{\rm collapse}  \right.
\\ \nonumber
 &+& \left. (s'-r') \frac{\phi(r',s')}{x_{13}^2 x_{24}^2 }   \;\;{\rm with}
\;\;
r' = \frac{x_{13}^2}{x_{12}^2}, s' = \frac{x_{23}^2}{x_{12}^2}
;\;4\rightarrow \infty \;\;{\rm collapse} \right)
\eea
It is now clear from 
\eq{dc1}, \eq{dc2} and \eq{dc3}, 
that after taking the limits indicated and using the 
fact $\phi(r,s)$ is a symmetric function in $r$ and $s$ we see that
the sum of the dangerous collapses among all the three body terms
cancel
\be{sumdan}
D(1;34) + D(3;41) + D(4;13) =0
\ee

This mechanism of cancellation of dangerous collapses cannot
hold when structure constant of interest is of a length conserving
process. This is because in a length conserving process the only 
genuine three body diagrams are when the two nearest neighbour
letters are on the longest operator  say on $O_\alpha$ and the
rest of the letters are on $O_\beta$ and $O_\gamma$. Therefore 
we cannot possibly have the last two terms in \eq{sumdan}. 
But, as we have mentioned in the previous section, in a length
conserving process there is a possibility of non-nearest neighbour
interactions which are planar. This is shown in \fig{leco}. 
\FIGURE{
\label{leco}
\centerline{\epsfxsize=16.truecm \epsfbox{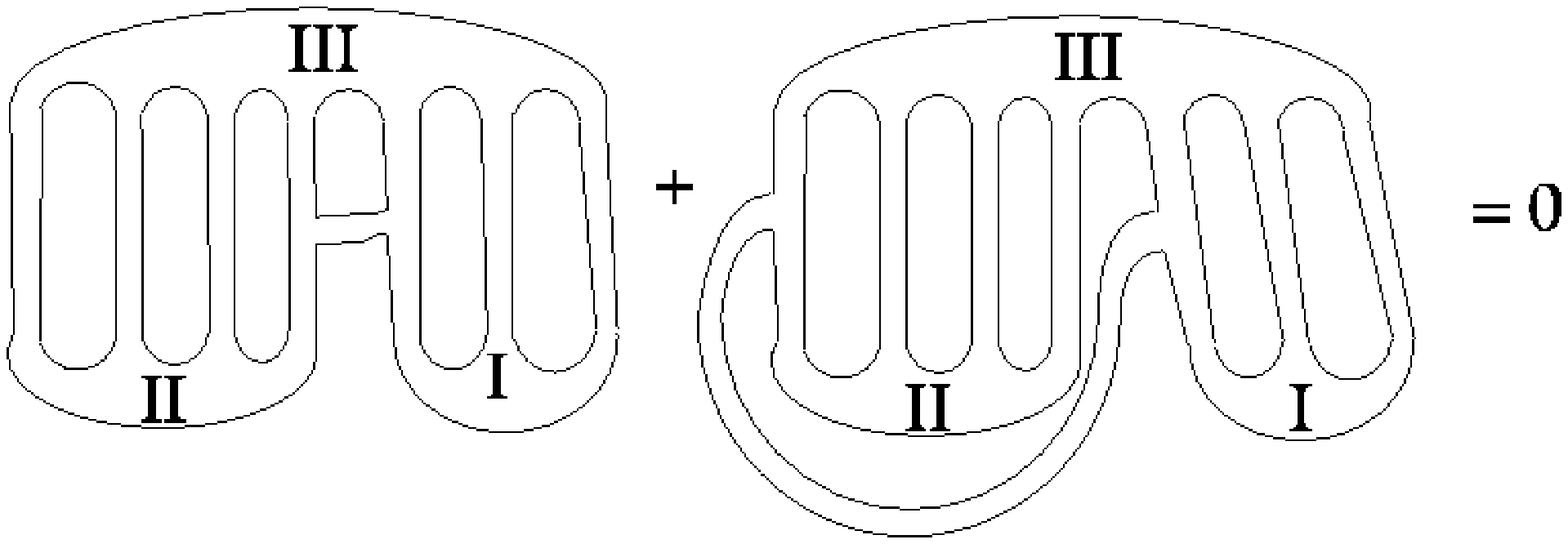}}
\caption{Cancellations in a length conserving process}
}
If one keeps track of the $U(N)$ group theoretical factors, it is
easy to show that there is a relative negative sign between the 
diagrams in \fig{leco}. Therefore such diagrams cancel, though
we will not go into details
in this paper, we have checked that for length
conserving process such diagrams ensure that 
the dangerous collapses in a length conserving process also cancel.

\subsection{An example}

In this subsection we consider a simple example to illustrate the
calculation of one loop corrections to structure constants.
We consider the following operators:
\be{defsopp}
O_\alpha = \frac{1}{\sqrt{N^3}} {\rm Tr} ( \phi^1\phi^2\phi^3) , 
\;\;\;\;
O_\beta = \frac{1}{\sqrt{N^3}} {\rm Tr} ( \phi^1\phi^2\phi^4) , 
\;\;\;\;
O_\gamma = \frac{1}{N} {\rm Tr} ( \phi^3\phi^4), 
\ee
the operators are at positions $x_1$, $x_3$  and $x_4$ respectively. 
The tree level correlation function of these operators are given by
\be{treescal}
\langle O_\alpha O_\beta O_\gamma \rangle ^{(0)} 
= \frac{1}{N} \frac{1}{x_{13}^4 x_{14}^2 x_{34}^2 }.
\ee
The one loop corrections will all have the above position dependent
factor multiplying the $\lambda$ dependent corrections. 
Below we write down the corrections from various diagrams, we divide
the contributions from genuine three body terms and two body terms. 
As we have seen in the previous section, we do not have to keep track
of the constants from the two body terms as they cancel in the metric
subtractions. Therefore we need to look at only the terms proportional 
to the logarithm in the
two body terms. The corrections to the structure constant will be
evaluated by \eq{s3pt}.

\vspace{.5cm}
\noindent
{ \emph{Three body terms} }

The three body terms consist of: 
\bea{3bodyscal}
&2& \left[  ( Q  + E + C_3 + C_4) (1;34) 
+  (Q + E + C_3 + C_4) (3;41) \right. \\ \nonumber
& +& \left.  (C_3 + C_4) (4;13)  \right],
\eea
here the labels $(1;34)$ refers to the diagram with two letters on 
the operator $O_\alpha$ and the remaining two letters on the operators
$O_\beta$ and $O_\gamma$ respectively. We have also suppressed the
$SO(6)$ index structure of each diagram for convenience, they can easily be
reinstated and evaluated.
Note that among the collapsed
diagrams we have written down only  the contributions of the
$3\rightarrow \infty$ and $4\rightarrow \infty$ collapse since
the remaining collapses are dangerous and cancel out.
For the diagrams of the type $(4;13)$ we have not written the quartic
term $Q$ and $E$, this is is because on examining the $SO(6)$
structure of these diagrams we see that they cancel among each other.
There is an overall factor of $2$ because of 
the presence of the outer three body diagrams.
We now give the  terms proportional to the logarithm 
of the above diagrams:
\bea{log3body}
&2& \left( -2 
\log\left( \frac{ x_{13}^2 x_{14}^2 }{x_{34}^2 \epsilon^2 } \right)  
+ \log \left( \frac{x_{13}^2}{\epsilon^2} \right)
+ \log \left( \frac{x_{14}^2}{\epsilon^2} \right)  \right. \\
\nonumber
&-&2 \log\left( \frac{ x_{34}^2 x_{13}^2 }{x_{14}^2 \epsilon^2 } \right)  
+ \log \left( \frac{x_{13}^2}{\epsilon^2} \right)
+ \log \left( \frac{x_{34}^2}{\epsilon^2} \right) \\ \nonumber
&+& \left. \log \left( \frac{x_{14}^2}{\epsilon^2} \right)
+ \log \left( \frac{x_{34}^2}{\epsilon^2} \right)  \right).
\eea
The logarithms in the above equation are the contributions of the respective
terms in \eq{3bodyscal}. 
Using \eq{s3pt}, 
the renormalization group invariant correction to the structure
constant is given by
\be{strcon}
 {\cal H}^{23}_{23} + {\cal H}^{24}_{24} + {\cal H}^{34}_{34} +{\cal
H}^{43}_{34} 
+ {\cal H}^{13}_{13} +{\cal H}^{14}_{14} +{\cal H}^{34}_{34} 
+{\cal H}^{43}_{13} = -8.
\ee
The indices on ${\cal H}$ refer to $SO(6)$ indices of the letters
involved.
Here the extra terms ${\cal H}^{43}_{34}$ is because of the fact that
the operator $O_\gamma$ is an operator of two letters whose position
can be interchanged.

\vspace{.5cm}

\noindent
{ \emph{Two body terms} }

As mentioned before, for the two body terms we have to focus only on
the log terms. The diagrams which contribute to this are:
\bea{2bodylog}
(Q + E + C_1 + C_2 + C_3 + C_4 ) (1;3) 
&+& 2 S(1;3) + S(1;4) + S(3;4),
\eea
where the labels $(1;3)$ indicate which two operators the
contributions arise from, we have again suppressed the $SO(6)$ indices
for convenience. Note that here all the $4$ collapses contribute, 
$S$ refers to the self energy contributions.
Evaluating these contributions we obtain
\bea{f2blog}
&-&2 \log \left( \frac{x_{13}^4}{\epsilon^4}\right) +  
4 \log \left( \frac{x_{13}^2 }{\epsilon^2 } \right)  
\\ \nonumber
&+& 
-8 \log \left( \frac{x_{13}^2 }{ \epsilon^2} \right) 
-4 \log \left( \frac{x_{14}^2 }{ \epsilon^2} \right) 
-4 \log \left( \frac{x_{34}^2 }{ \epsilon^2} \right). 
\eea

Combining \eq{log3body}, and \eq{f2blog} and \eq{strcon} we find that 
the log correction and the renormalization group invariant one loop
correction to the structure constant is given by
\be{allscalcor}
\frac{\lambda}{N} \left( -12 \log\left(\frac{x_{13}^2}{\epsilon^2} \right)  -8
\right).
\ee
Here we have reinstated the factor $\lambda/N$ which occurs in the
corrections to the structure constants.

\section{ Operators with derivatives}

In the previous section we showed that the anomalous dimension
Hamiltonian controls the corrections to structure constants in the
$SO(6)$ sector. There were basically three reasons for this: 
(i) the $SO(6)$ spin dependent term factorizes out in the calculations,
(ii) ${\cal N}=4$ supersymmetry ensures that quartic term and the 
gauge exchange terms comes with the same coupling constant, 
(iii) contributions of all collapsed diagrams canceled.
As we have argued in the introduction,  since ${\cal N}=4$ super
Yang-Mills admits a string dual, the structure constants of 
the theory should be determined 
basically by the geometric delta function overlap
of the dual string theory. One can see that at 
$\lambda =0$ and at large $N$ ensures that three point functions of 
single trace gauge invariant operators can be written as delta
function overlap in a string bit theory \cite{adgn:2005}. 
Turning on finite $\lambda$ renders $\alpha'$  of 
the string theory finite, and induces
nearest neighbour interactions between the bits. Thus, the 
modifications to structure constants must be 
only due to effects of 
interaction in the propagation of the bits, the geometric 
delta function overlap of the string is invariant.
The fact that in the $SO(6)$ sector the one loop corrections to 
the structure constants 
is dictated by the anomalous dimension
Hamiltonian indicates the possibility that it is only the 
world sheet Hamiltonian in the bit string theory 
which is necessary to compute corrections to structure
constants. 
To verify this and to 
identify the precise operator 
which is responsible for the propagation of 
the bits we need to compute one loop corrections to structure
constants
with more general operators outside the $SO(6)$ scalar sector.
Among the three simplifications in the $SO(6)$ sector discussed above, 
the  factorization of $SO(6)$ spin dependent term 
will not be present if there are
derivatives in the letters. 
This motivates the  evaluation of  one loop corrections to structure
constants of operators with derivatives.

\subsection{Primaries with derivatives}

Before we start the one loop calculation, we need to specify the 
operators with derivatives which are conformal primaries that
we will be dealing with.
We work with operators having $SO(6)$ scalars with arbitrary number of
derivatives in  a fixed complex direction.
For example the following
operator 
\be{exop}
{\rm Tr} ( D^{m_1}_z \phi^{i_1} D_z^{m_2} \phi^{i_2}\cdots   \cdot D^{m_j}_z
\phi^{i_j} \cdots ),
\ee
where $D_z = \partial_z + i g
[A_z, \;\cdot \;]$ \footnote{In our notation $g^2 =
\frac{g_{YM}^2}{2(2\pi)^2}$, see appendix A. }
is the covariant derivative in a given complex
direction $z= x^2 + i x^3$, $m_j$ refers to the number of
derivatives on the $j^{{\rm th}}$ letter. To construct the primaries 
at tree level we can ignore the commutator term in the covariant
derivative.  
To construct a conformal primary from such operators we need to know
the action of the special conformal transformations
$K_\mu$ on these states. The action of $K_\mu$ on a scalar is given by
\be{actk}
[K_\mu, \phi] = (2x_\mu x\cdot \partial     +  2x_\mu - x^2
\partial_\mu)
\phi.
\ee
Since all the fields are at the origin and the derivatives are only in
the holomorphic direction we can set all other coordinates in $K_z$ to
zero, this gives 
\be{kz}
K^z = z^2\partial_z + z,
\ee
similarly the other generators are given by
\be{othsl2}
P_z = \partial_z, \;\;\; D = 1+
z\partial_z.
\ee
They satisfy the algebra
\be{kalg}
[D, K^z] = K^z, \;\;\; [D, P_z] = -P_z \;\;\; [P_z, K^z] = 
2z \partial_z +1  = D + M_{z\bar{z} }
\ee
where $M_{z\bar{z}} = z\partial_z$ is the angular momentum generator
on the $z$ plane when $\bar{z}$ is set to zero. 
The above algebra forms an $SL(2)$ algebra, to see
this identify 
\be{sl2}
J_3 = -\frac{1}{2} \left( D+ M_{z\bar{z}} \right)  , \;\;\;\;\; 
J_+ = P_z,\;\;\;\;\;\; J_- = K^z,
\ee
then we have
\be{sl2alg}
[J_3, J_{\pm} ] = \pm J_{\pm}, \;\;\;\;  [J_+, J_-] = -2 J_3.
\ee
Thus scalars with derivatives in  a given holomorphic sector form
representations of the $SL(2)$ algebra.
The action of  $K_z$  a scalar with $m$  derivatives is given by
\be{act}
[K_z,  \frac{\partial^m}{m!} \phi^i] = 
m\frac{1}{(m-1)!} \partial^{m-1}\phi^i.
\ee
Here we have divided the $m$th derivative by $m!$ to ensure the 
two point function of these derivatives are normalized to $1$, 
we have also suppressed the subscript $z$ on the derivatives which 
will be understood for the rest of paper.
It is easy to construct primaries by suitably taking  linear
combinations of these operators. For example a simple class of
primaries with derivatives only on two of the scalars is given by
\be{priclas}
\sum_{m=0}^n  (-1)^m {\;}^{n}C_m {\rm Tr} \left(
 \frac{ \partial^m \phi^{i_1}}{m!} \phi^{i_2} \cdots  
\frac{ \partial^{n-m} \phi^{i_j} }{(n-m)!} \phi^{i_{j+1} } \cdots 
\right).
\ee
Similarly,  combinations of operators with derivatives only in
the anti-holomorphic direction $\bar{z}$ can be chosen so that
they are primaries.

Three point functions as well as two point functions
of primaries have definite tensor structure 
as given in \eq{3ptdef} and \eq{g2pt} respectively.
Therefore it is sufficient to focus
terms proportional to products of 
of the identity $\delta_{\mu\nu}$ in the tensor structure.
For operators with derivatives only in the holomorphic or the
anti-holomorphic direction it is sufficient to look at terms
proportional to products of the identity $\delta_{z\bar{z}}  $. 
This simplifies calculations considerably: for instance in the
calculation of the interaction  with 4 letters, the number of
holomorphic derivatives must equal the number of anti-holomorphic
derivatives. 
Finally, another useful fact about the $SL(2)$ sector is that 
when the scalars are in a given Cartan direction of $SO(6)$, the
detailed calculation of the 
the anomalous dimension Hamiltonian has been done in
\cite{Beisert:2003jj}.

\subsection{The processes}

From the slicing argument and our detailed discussion for the $SO(6)$
sector, the 
corrections to the structure constants are governed by the constants
in the following basic quantity
\bea{dbasic}
& & \left( U^{(i_a,m_a) ( i_{a+1}, m_{a+1}) }
_{(j_{b+1}, n_{b+1}) (k_c, s_c) } ({\rm 3pt}) - 
\frac{1}{2} U^{(i_a , m_a) (i_{a+1}, m_{a+1}) }
_{(j_{b+1}, n_{b+1})( k_c, s_c)} ({\rm 2pt})\right)
\delta^{j_b}_{k_{c+1}}\delta(n_b, s_{c+1}) \\ \nonumber
&+&
\left( U^{(j_b, n_b) ( j_{b+1}, n_{b+1})  }
_{(k_{c+1}, s_{c+1})  (i_a, m_a) } ({\rm 3pt}) - 
\frac{1}{2} U^{(j_b, n_b) (j_{b+1}, n_{b+1}) }
_{(k_{c+1}, s_{c+1})(i_a, m_a)} ({\rm 2pt})\right)
\delta^{k_c}_{i_{a+1}} \delta(s_c, m_{a+1}) \\ \nonumber
&+&
\left( U^{(k_c, s_c)(k_{c+1}, s_{c+1}) }
_{(i_{a+1}, m_{a+1})( j_b, n_b)} ({\rm 3pt}) - 
\frac{1}{2} 
U^{(k_c, s_c)(k_{c+1}, s_{c+1}) }
_{(i_{a+1}, m_{a+1})( j_b, n_b)} ({\rm 2pt})  \right)
\delta^{i_a}_{j_{b+1}} \delta (m_a, n_{b+1}).
\eea
In the above formula $i, j, k$ label $SO(6)$ indices and  $m, n, s$
label the number of derivatives which could be either holomorphic or
anti-holomorphic.
$a, b, c$ refers to the position of
the letters in each of the operators. $\delta(m , n)$ refers to the
delta function which is one when either the number of holomorphic
$m$ equals the number of anti-holomorphic derivatives $n$ or vice
versa. 
To further simplify our analysis we will restrict our attention
to the cases  when the total number of holomorphic derivatives on the
operator with $2$ letters adjacent to each other
in the interaction, is always greater
that the number of anti-holomorphic derivatives on either of the
letters of the remaining two operators. But, the methods developed
here can be applied to study the other cases also.
Let us work with only holomorphic derivatives on $O_\alpha$ and 
anti-holomorphic derivatives on $O_\beta$ and $O_\gamma$. Then, our
restriction implies that for the first term in \eq{dbasic}
$m_a +m_{a+1} \geq n_{b+1}, s_c$.

We now detail all the processes involved in the evaluation of  the
constants in the interaction
$U^{(i, m)(j n)}_{(k, s)(l, t)}$.
We again use the point splitting scheme to evaluate the diagrams. 
For the 3pt contribution 
the letters $D^m \phi^i/m!$ and $D^n \phi^j/n!$ are at positions
$x_1$ and $x_2$ respectively such that $x_2 - x_1=\epsilon$ with $\epsilon
\rightarrow 0$ and the letters $\bar{D}^s \phi^k/s!$ and $\bar{D}^t
\phi^l/t!$ are at $x_3$ and $x_4$ respectively. 
For the 2pt contribution one further takes the  limit
$x_4 \rightarrow x_3 =\epsilon$. In all the diagrams we will first
perform the relevant derivatives and then take the appropriate limits.
Since we are looking for only the term proportional to the identity
we have the constraint $m + n = s + t $, the number of
holomorphic derivatives must be equal to the number of 
anti-holomorphic derivatives. 
\vspace{.5cm}

\noindent
{\emph{ (i) The quartic interaction}}

The contribution of the quartic interaction of scalars to 
$U^{(i, m)(j ,n)}_{(k, s)(l, t)}$ is shown in the \fig{gaexc}. 
\FIGURE{
\label{gaexc}
\centerline{\epsfxsize=16.truecm \epsfbox{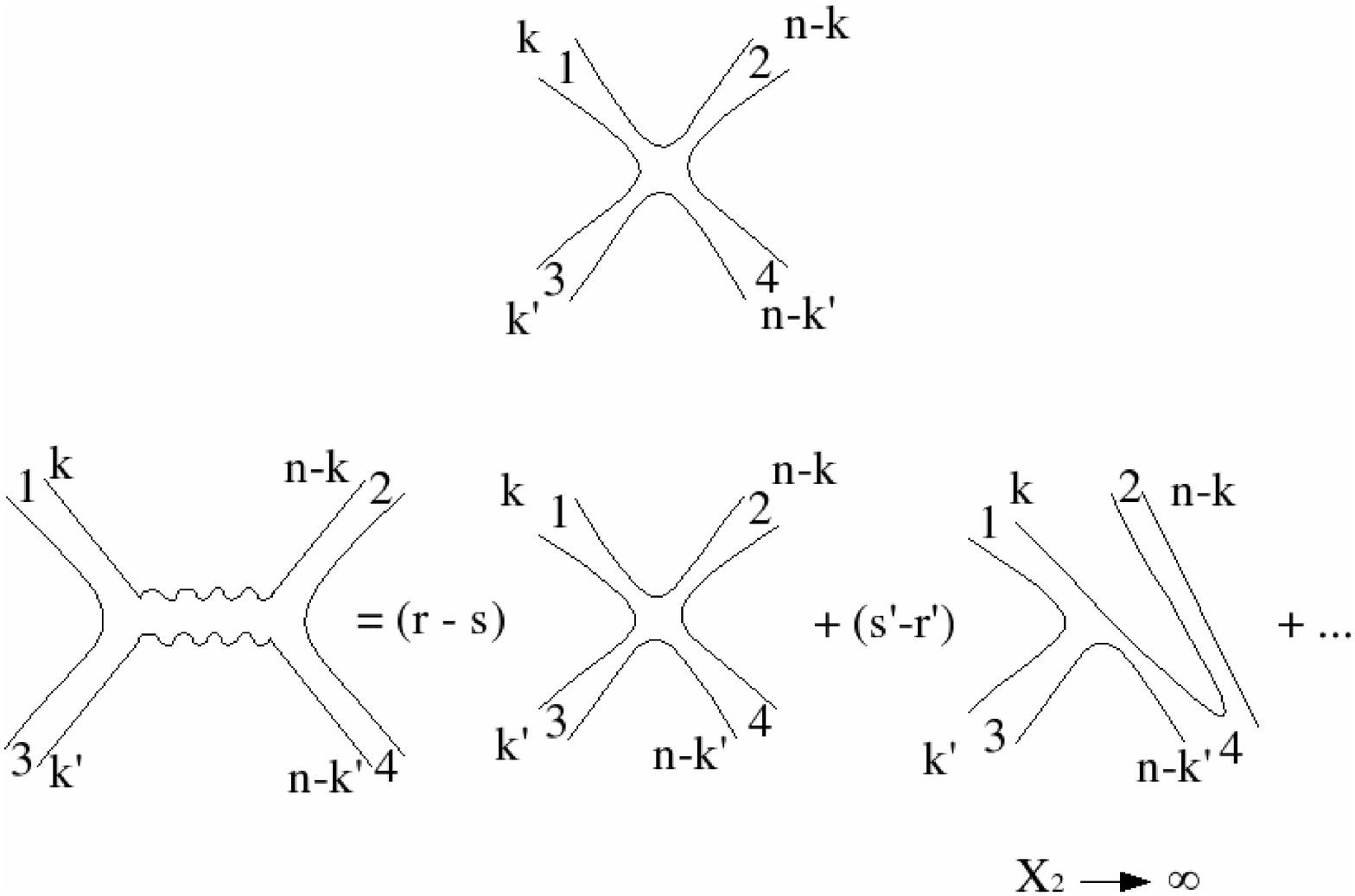}}
\caption{The quartic and the gauge exchange with
$x_2\rightarrow\infty$ collapse }
}
We first focus on the $3$ pt contribution: 
the constant and the log part of this interaction 
can be extracted by  evaluating the
limits in 
\be{dquart}
Q^{(i,m), (j,n)}_{(k,s)(l,t)}({\rm 3 pt} ) =
 \left( 2 \delta^{j}_k \delta^i_l -
\delta^i_k \delta^j_l - \delta^{ij}\delta_{kl} \right)  
\lim_{x_2 \rightarrow x_1}
\frac{\partial_1^m \partial_2^n \bar\partial_3^s \bar\partial_4^t}{
m! n!s! t!}
\left( \frac{\phi(r,s) }{x_{13}^2 x_{24}^2 }  \right).
\ee
Now one can use the expansions of $\phi(r,s)$ in 
\eq{usexp} and perform the appropriate derivatives.
In the above equation $\partial_1$ 
and $\partial_2$ refers to the holomorphic
derivative in the $z_1$ and $z_2$ direction respectively, while
$\bar\partial_3$ and $\bar\partial_4$ refers to the 
anti-holomorphic derivative in the $\bar{z}_1$ and $\bar{z}_2$
directions respectively.
Taking the derivatives is sufficiently simple as one has to focus only
on the term proportional to the 
identity $\delta_{z\bar{z}}$ 
since we are  dealing  with primaries, finally one has
to take the limit $x_2\rightarrow x_1$.
The general form of the quartic term is given by
\be{genquart}
Q^{(i,m), (j,n)}_{(k,s)(l,t)}({\rm 3 pt} ) =
 \left( 2 \delta^{j}_k \delta^i_l -
\delta^i_k \delta^j_l - \delta^{ij}\delta_{kl} \right)  
\frac{1}{x_{13}^{2( s+1) } x_{14}^{2(t+1)} } \left(
{\cal A}_Q  
\log \left(\frac{ x_{13}^2 x_{14}^2 } { x_{34}^2 \epsilon^2} \right) 
+  {\cal C}_Q \right).
\ee
The coefficient of the log  ${\cal A}_Q$ 
and the constant ${\cal C}_Q$ for the various cases 
can be read from 
table 3.  of  appendix C. 
The quartic interaction contribution to the corresponding
2pt term is given by further taking the limit $x_4\rightarrow x_3$, 
thus the constant obtained for the 2pt term will be  the same as constants
of the 3pt term. 
\vspace{.5cm}

\noindent
{\emph{ (ii) Gauge exchange}}

The gauge exchange contribution to $U (3 {\rm pt} )$ 
can be found by evaluating the limit in 
\bea{dgex}
G^{(i,m), (j,n)}_{(k,s)(l,t)}({\rm 3 pt} ) &=& 
\delta^i_k \delta^j_l \lim_{x_2\rightarrow x_1} 
\frac{\partial_1^m \partial_2^n \bar\partial_3^s \bar\partial_4^t
}{ m!n!s!t!} H, \\
\nonumber
&=& \delta^i_k \delta^j_l \lim_{x_2\rightarrow x_1} 
\frac{\partial_1^m \partial_2^n \bar\partial_3^s \bar\partial_4^t }{
m!n!s!t!}
\left( E + C_1 + C_2 + C_3 + C_4 \right),
\eea
where 
\be{defec}
E = ( r- s) \frac{\phi(r,s)}{x_{13}^2 x_{24}^2 },
\ee
and $C_1, C_2, C_3, C_4$ are the 
collapsed diagrams given in \eq{heq}. In \eq{dgex} we have basically used the
\eq{heq} to write the gauge exchange diagram in terms of the various
collapses and \eq{defec}. The equation \eq{heq} is  true
when all the points $x_1, x_2, x_3, x_4$ are strictly distinct.
Therefore, we use the equation when all the points are distinct, 
take the appropriate derivatives and then finally take the limit
$x_2 \rightarrow x_1$. 
Just as the quartic diagram,
the general form for  the diagram $E(3{\rm pt})$  is given by
\be{genex}
E ( 3{\rm pt} ) = \delta^i_k \delta^j_l 
\frac{1}{x_{13}^{2( s+1) } x_{14}^{2(t+1)} } \left(
{\cal A}_E  
\log \left(\frac{ x_{13}^2 x_{14}^2 } { x_{34}^2 \epsilon^2} \right) 
+  {\cal C}_E \right).
\ee
In tables 4. and 5 of 
appendix C. we tabulate the values of 
${\cal A}_E$ and ${\cal C}_E$
for the various cases.

We now 
examine the structure of the derivatives in each of the
collapses and  list the
conditions under which they contribute to the identity.
Consider the $1\rightarrow\infty$ collapse, which is given by
\bea{c1}
C_1 &= &
\delta^i_k \delta^j_l \lim_{x_2\rightarrow x_1} 
\frac{\partial_1^m \partial_2^n \bar\partial_3^s \bar\partial_4^t }{
m!n!s!t!}\left(  (r'-s')  \frac{\phi(r', s')}{x_{13}^2 x_{24}^2}
\right),
\\ \nonumber
& & {\rm with} \;\;\;\;\; r' = \frac{x_{34}^2}{x_{24}^2} 
, \;\;\;\; s' = \frac{x_{23}^2 }{ x_{24}^2}.
\eea
Note that if $m>s$ and therefore $n<t$, there is no possibility of
saturating the derivatives in the $z_1$ direction to give 
a term proportional to the identity, since $r'$ and
$s'$ are independent of $x_1$. Therefore, this collapse diagram
contributes to terms proportional to the identity only when 
$m\leq s$ and therefore $n\geq t$. A similar analysis with all the
collapses leads to the following table: 
\vspace{.5cm}
\begin{center}
\begin{tabular}{l | l | l|  l }
Diagram & $m>s;\;\; t>n$ & $m <s; \;\;t<n$ & $m=s; \;\; n=t$ \\ \hline
$C_1$ & No & Yes & Yes \\ \hline
$C_2$ & Yes & No& Yes \\ \hline
$C_3$ & Yes & No & Yes \\ \hline
$C_4$ & No & Yes & Yes \\ \hline
\end{tabular}
\\
\vspace{.5cm}
{\bf\small {Table 1.}} Conditions for the contribution of the
collapsed diagrams.
\end{center}
It details the conditions
on $m, n,s,t$ under which various collapse diagrams contribute to the
term proportional to the identity.

Just as in the case of the 
$SO(6)$ sector discussed in
the previous section, the collapses $C_1$ and $C_2$ are
potentially dangerous as the values of $r'$ and $s'$ for these
collapses do not tend to either $0$ and $1$ respectively under the
limit $x_2 \rightarrow x_1$. 
Therefore, $C_1$ and $C_2$ 
are non trivial functions not just logarithms  or constants 
which are required by conformal invariance.  
As discussed in the previous section for the 
$SO(6)$ sector, these potentially dangerous
collapses must cancel out leaving behind only logarithms or constants. 
The detailed mechanisms which are responsible for this 
in this sector will be
discussed in the next subsection.

For the evaluation of 
$G^{(i,m), (j,n)}_{(k,s)(l,t)}({\rm 2 pt} )$ we have to also take 
$x_4\rightarrow x_3$ limit in addition to the $x_2\rightarrow x_1$
limit. On taking both these limits 
it is easy to see that $r'$ and $s'$ for the
$1\rightarrow \infty$ and $2\rightarrow\infty$ collapse 
also tend to $0$ and $1$ respectively. 
Therefore all the collapses reduce to 
logs and constants. 
\vspace{.5cm}

\noindent
{ \emph{ (iii) Gauge bosons on one external leg} }

The covariant derivatives on the letters also have gauge bosons, at
one loop one such external gauge boson from say $D^{m}\phi^i$ can
interact with the letters $D^n \phi^j$, $D^t\phi^l$ 
as show in \fig{onextg}.  
\FIGURE{
\label{onextg}
\centerline{\epsfxsize=16.truecm \epsfbox{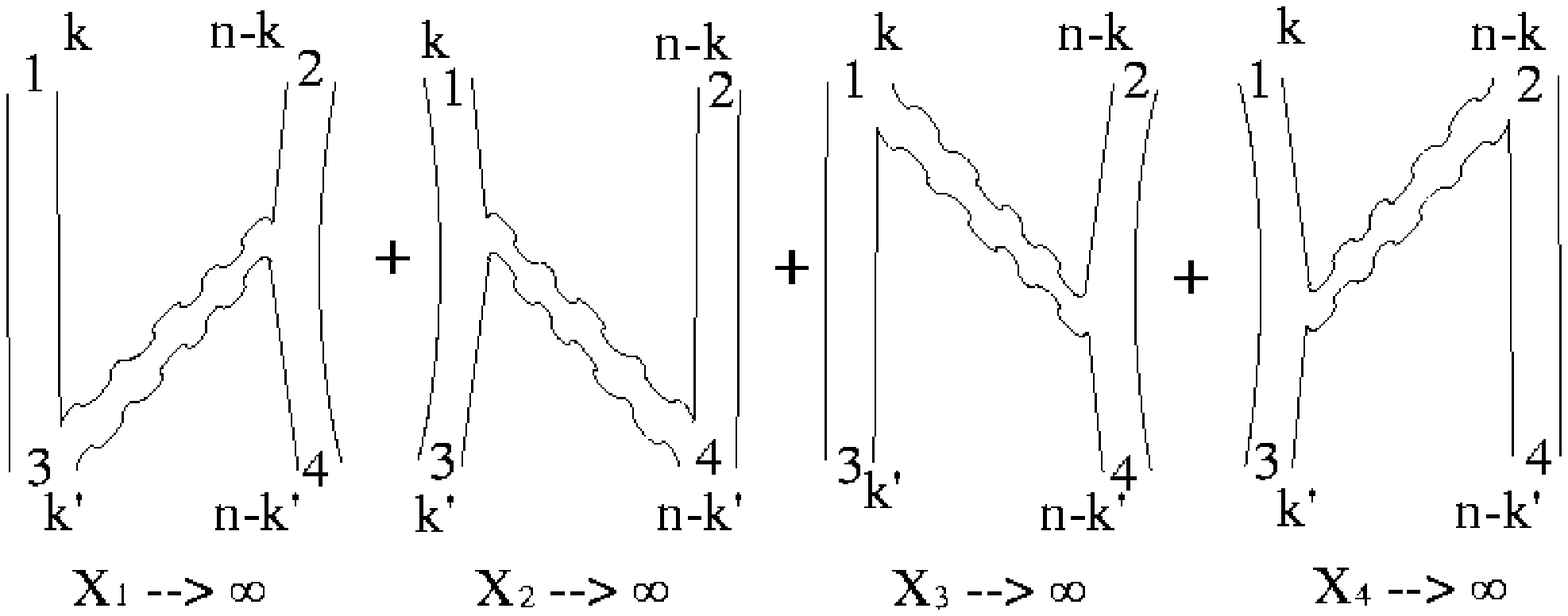}}
\caption{Diagrams with gauge boson on one external leg.}
}
To evaluate this diagram it is convenient to expand the covariant
derivative
to  order one in the $g_{YM}$ as:
\be{defcovder}
D^m \phi = \partial^m \phi + i g \sum_{p=1}^m  {\;}^{m}C_p [
\partial^{m-1}A_z, \partial^{m-p} \phi].
\ee
Other similar process with one external gauge boson on the other 3
letters exist,
these are shown in \fig{onextg}. 
We now write the interaction term of each
such diagram.  The contribution of the diagram with the gauge boson on
the letter $D^m \phi^i$ is given by
\bea{dexg1}
A_3(3{\rm pt} )  
&=&\delta^{i}_{k} \delta^j_l 
\frac{1}{ m!n!s!t!} \times \\ \nonumber
& &\lim_{x_2\rightarrow x_1} 
\sum_{p=1}^m  {\;}^mC_p 
\left( \partial_1^{m-p} \bar\partial_3^s \frac{1}{x_{13}^2}  \right)
\left( \partial_1^{p-1} ( 2\partial_2 + \partial_1 ) \partial_2^n
\bar\partial_4^t
\frac{\phi(r',s')}{x_{24}^2}\right) ,  \\ \nonumber
&\;&\; {\rm where} \;\;\;r' = \frac{x_{12}^2 }{x_{24}^2}, \;
s' = \frac{x_{14}^2}{x_{24}^2}.
\eea
We have labelled this diagram $A_3$ as the values of $r'$ and $s'$
that occur are the values of the $3\rightarrow \infty$ collapse. Note
that we have used momentum conservation on the vertex of a gauge boson
with two scalars. From the structure of the derivatives in the first
bracket of \eq{dexg1}, it is clear the term proportional to identity
occurs only when $m>s$.
Similarly the diagram with the external gauge boson on the letter
$D^n \phi^j$ is given by
\bea{dexg2}
A_4(3{\rm pt} ) 
&=&\delta^{i}_{k} \delta^j_l  
\frac{1}{ m!n!s!t!} \times \\ \nonumber
& &\lim_{x_2\rightarrow x_1} 
\sum_{p=1}^n  {\;}^nC_p 
\left( \partial_2^{n-p} \bar\partial_4^t \frac{1}{x_{24}^2}  \right)
\left( \partial_2^{p-1} ( 2\partial_1 + \partial_2 ) \partial_1^m
\bar\partial_3^s
\frac{\phi(r',s')}{x_{13}^2}\right) ,  \\ \nonumber
&\;&\; {\rm where} \;\;\;r' = \frac{x_{12}^2 }{x_{13}^2}, \;
s' = \frac{x_{23}^2}{x_{13}^2}.
\eea
This diagram contributes to terms proportional to the identity only
when $n>t$.
If the external gauge boson is from the letter $D^s \phi^k$ 
the interaction is given by
\bea{dexg3}
A_1(3{\rm pt} ) 
&=&\delta^{i}_{k} \delta^j_l 
\frac{1}{ m!n!s!t!} \times \\ \nonumber
& &\lim_{x_2\rightarrow x_1} 
\sum_{p=1}^s  {\;}^sC_p 
\left( \bar\partial_3^{s-p} \partial_1^m \frac{1}{x_{13}^2}  \right)
\left( \bar\partial_3^{p-1} 
( 2\bar\partial_4 + \bar\partial_3 ) \partial_2^n
\bar\partial_4^t
\frac{\phi(r',s')}{x_{24}^2}\right) ,  \\ \nonumber
&\;&\; {\rm where} \;\;\;r' = \frac{x_{34}^2 }{x_{24}^2}, \;
s' = \frac{x_{23}^2}{x_{24}^2}.
\eea
Here the above diagram contributes only when $s>m$.
Finally when the external gauge boson is from the  letter
$D^t\phi^l$, the diagram is given by
\bea{dexg4}
A_2(3{\rm pt} ) 
&=&\delta^{i}_{k} \delta^j_l 
\frac{1}{ m!n!s!t!} \times \\ \nonumber
& &\lim_{x_2\rightarrow x_1} 
\frac{1}{ m!n!s!t!}
\sum_{p=1}^t  {\;}^tC_p 
\left( \bar\partial_4^{t-p} \partial_2^n \frac{1}{x_{24}^2}  \right)
\left( \bar\partial_4^{p-1} 
( 2\bar\partial_3 + \bar\partial_4 ) \partial_1^m
\bar\partial_3^s
\frac{\phi(r',s')}{x_{13}^2}\right) ,  \\ \nonumber
&\;&\; {\rm where} \;\;\;r' = \frac{x_{34}^2 }{x_{13}^2}, \;
s' = \frac{x_{14}^2}{x_{13}^2}.
\eea
This contributes only when $t>n$. We summarize  the conditions
on $m, n, s, t$ under which all these diagrams contribute to the 
term proportional to identity in the following table:
\vspace{.5cm}
\begin{center}
\begin{tabular}{l | l | l|  l }
Diagram & $m>s;\;\; t<n$ & $m <s; \;\;t<n$ & $m=s; \;\; n=t$ \\ \hline
$A_1$ & No & Yes & No \\ \hline
$A_2$ & Yes & No& No \\ \hline
$A_3$ & Yes & No & No \\ \hline
$A_4$ & No & Yes & No \\ \hline
\end{tabular}
\\
\vspace{.5cm}
{\bf\small {Table 2.}} Contributions of diagrams
with gauge boson on one leg.
\end{center}

Note that the external gauge boson contribution 
$A_1$ and $A_2$ given in \eq{dexg3} and
\eq{dexg4} respectively 
are non trivial functions of the respective $r'$ and $s'$,
as these do not reduce to either logarithms or constants under the
limit $x_2\rightarrow x_1$. Therefore
contributions from these diagrams can potentially violate conformal
invariance. But, we will show that  contributions from these
terms add up with the 
dangerous collapses $C_1$ and $C_2$ of \eq{dgex}
to finally give only logarithms and constants ensuring conformal
invariance. 
As an indication of this we see  that from table 2. and table 1. that
whenever $A_1$ or $A_2$ contributes to the term proportional to the
constant $C_1$ or $C_2$ also contributes. 
The mechanism of how this comes about will be discussed
in detail in the next subsection.
\vspace{.5cm}

\noindent
{ \emph{ (iv) Gauge bosons on two legs} }

Diagrams with  gauge bosons on two different legs contribute constants 
at one loop. These diagrams are just planar Wick contractions with
the gauge bosons on the respective external legs. 
The ones which contribute to $U$ are the first two diagrams of
\fig{gbtl}. 
\FIGURE{
\label{gbtl}
\centerline{\epsfxsize=16.truecm \epsfbox{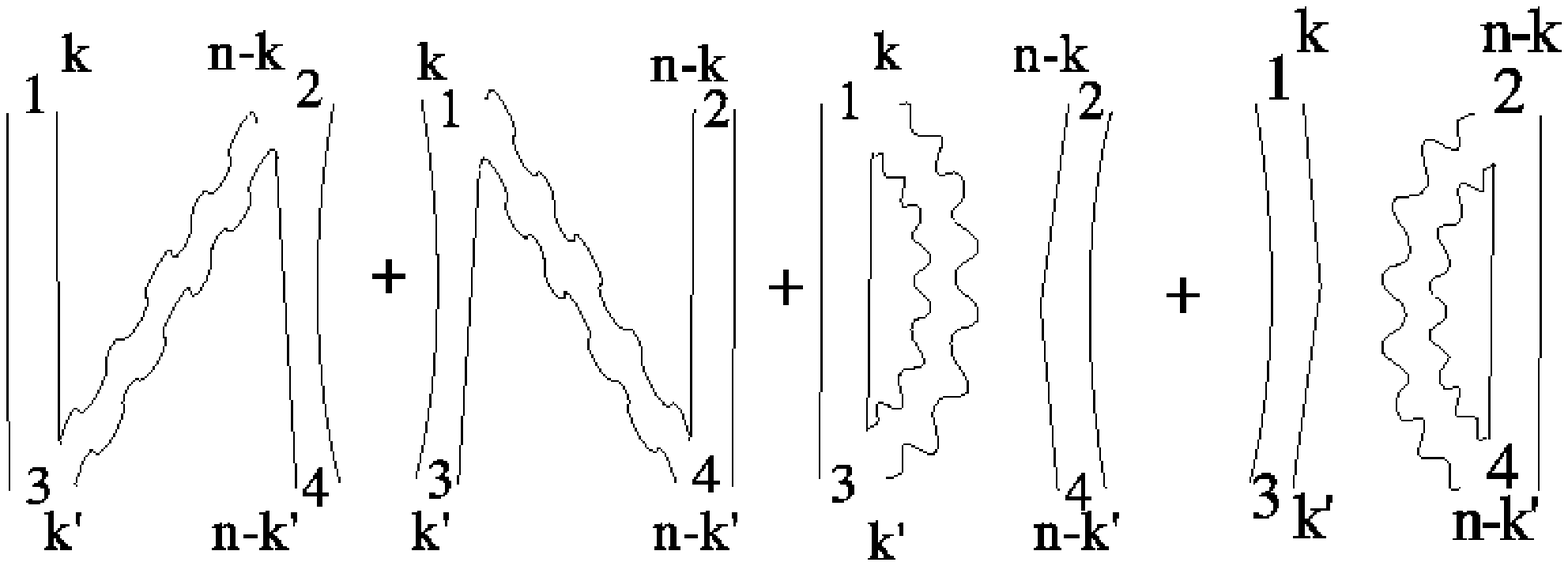}}
\caption{Gauge bosons on two legs}
}
The ones with
the external gauge boson from the letter $D^m \phi^i$ and $\bar{D}^t\phi^l$
is given by
\bea{d2g1}
B_1 &=& - 2\delta^i_k \delta^j_l \frac{1}{m!n!s!t!}  \times \\
\nonumber
& & \sum_{p=1}^m \sum_{p'=1}^t {\;}^mC_p
{\;}^tC_{p'} \bar\partial_3^s \partial_1^{m-p} 
\left( \frac{1}{x_{13}^2}  \right) \partial_1^{p-1}
\bar\partial_4^{p'-1} \left( \frac{1}{x_{14}^2} \right)
\partial_2^n \bar\partial_4^{t-p'} \frac{1}{x_{24}^2}.
\eea
The presence of the negative sign in the above formula is due to the
fact that the gauge fields on the two legs 
come on two different sides of the
commutator. The factor of $2$ occurs in \eq{d2g1} if one keeps track
the factors of $2$ in $g^2$  and uses the fact that
\be{propaz}
\langle A_z^a (x_1) A_{\bar z}^a (x_2) \rangle = 
\delta^{ab} \frac{1}{2 (x_1 -x_1)^2 }.
\ee
Looking for the term proportional to the identity, we see that
the above  diagram contributes only when $m >s$ and therefore
$n<t$, evaluating the constant we obtain
\be{d2g1c}
B_1 = - 2\delta^i_k \delta^j_l\frac{1}{(m -s)^2},
\ee
where we have used  $m+n=s+t =q$. 
Similarly the contribution 
with the external gauge boson from the
letter $D^n\phi^j$ and $\bar{D}^s\phi^k$  is given by
\bea{d2g2}
B_2 &=& -2\delta^i_k\delta^j_l \frac{1}{m!n!s!t!} \times \\ \nonumber
\lim_{x_2\rightarrow x_1}
& &\sum_{p=1}^s \sum_{p'=1}^n {\;}^sC_p
{\;}^nC_{p'} \bar\partial_3^{s-p} \partial_1^{m} 
\left( \frac{1}{x_{13}^2}  \right) \bar\partial_3^{p-1}
\partial_2^{p'-1} \left( \frac{1}{x_{23}^2} \right)
\partial_2^{n-p'} \bar\partial_4^{t} \frac{1}{x_{24}^2}.
\eea
Again looking for the term proportional to the identity we see that
the above term contributes only when $s>m$ and $n>t$. Keeping track of
the constant term we see that it is given by
\be{d2g22}
B_2 = - 2\delta^i_k\delta^j_l\frac{1}{(s-m)^2}.
\ee
Note that both these diagrams do not contribute if $m=s$ or $n=t$.

Consider the remaining contributions from the gauge boson on two legs
(see \fig{gbtl}.), 
for instance the diagram with the external gauge boson from the
letter $D^m\phi^i$ and $D^s\phi^k$. These diagrams are 
two body terms and their contribution 
to the renormalization scheme independent corrections to the
three point functions cancel by the slicing argument.

\subsection{Mechanisms ensuring conformal invariance}

\vspace{.5cm}
\noindent
{\emph{ Case 1. $m>s$; $t>n$ } }

From table 1. and table 2.  it is clear that only 
the collapsed diagram $C_2$ and 
the external gauge boson on one leg 
$A_2$ are the potentially dangerous diagrams which can
violate conformal invariance for this case.
We show that both these diagrams combine in 
a non-trivial way to give only logarithms or constants.
To simplify matters we first discuss the case of $m=1, s=0, n=0, t
=1$, then $C_2$ is given by
\bea{mcic2}
C_2 &=& \delta^{i}_k \delta^j_l \partial_1\bar\partial_4 \left(
\frac{1}{x_{13}^2 x_{24}^2} ( s'-r') \phi(r', s') \right), 
\;\;\;\;\; r' =\frac{x_{34}^2}{x_{13}^2}, \;\; s'
=\frac{x_{14}^2}{x_{13}^2}, \\ \nonumber
&=& \delta^i_k \delta^j_l \frac{1}{x_{13}^4 x_{24}^2 }
\left[ - \phi -  (s'-r')\partial_{s'} \phi \right],
\eea
here, in writing the second line 
we have kept only the terms proportional to the identity while
performing the differentiation.
The contribution of $A_2$ can be read out from \eq{dexg4}, it is given by
\bea{mcia2}
A_2 &=& \delta^i_k\delta^j_l
\frac{1}{x_{24}^2} \left[  ( 2 \bar\partial_3 \partial_1 +
\bar\partial_4 \partial_1 ) \frac{\phi(r',s')}{x_{13}^2} \right], \\
\nonumber
&=& \delta^i_k\delta^j_l
\frac{1}{x_{24}^2 x_{13}^4} \left[
2 \phi + 2 ( r'\partial_{r'} + s'\partial_{s'} ) \phi
-\partial_{s'} \phi \right].
\eea
Adding $C_2$ and $A_2$  form \eq{mcic2} and \eq{mcia2} we obtain
\be{sumci}
C_2 + A_2 = \delta^i_k \delta^j_l \frac{1}{x_{24}^2 x_{13}^4 } 
\left( \phi + ( r'+s'-1) \partial_{s'} \phi 
+ 2 r' \partial_{r'} \phi \right) .
\ee
Note that on adding $C_2$ and $A_2$, the combination of $\phi(r',s')$
in the bracket of the above equation
is precisely that of  \eq{de}. In appendix B. it is shown that
$\phi(r',s')$ satisfies the 
inhomogeneous partial differential equation
\be{deaux}
 \phi + ( r'+s'-1) \partial_{s'} \phi 
+ 2 r' \partial_{r'} \phi  = - \frac{\log r'}{s'}.
\ee
The differential equation ensures that though $\phi(r',s')$ is a
nontrivial function of $r'$ and $s'$ not just logarithms or constants, the
combination which occurs in $A_2$ and $C_2$ is such that it reduces to
a logarithm  ensuring conformal invariance. 
Substituting this in \eq{sumci} we obtain
\be{sumci2}
C_2 + A_2 =
\delta^i_k \delta^j_l \frac{1}{x_{24}^2 x_{13}^2 x_{14}^2 } \ln \left( \frac{
x_{13}^2}{x_{34}^2} \right). 
\ee
Now it is also clear that one needs the additional $1/s'$ on the right
hand side  
of\eq{deaux} to obtain the right powers of $x$ dictated by conformal
invariance.
Finally taking the limit $x_2\rightarrow x_1$ we obtain
\be{sumci3}
C_2 + A_2 = \delta^i_k\delta^j_l\frac{1}{x_{14}^4 x_{13}^2 } \log \left(
\frac{ x_{13}^2}{x_{34}^2} \right).
\ee
We have illustrated this mechanism of ensuring conformal invariance
in \fig{diffpi}
\FIGURE{
\label{diffpi}
\centerline{\epsfxsize=16.truecm \epsfbox{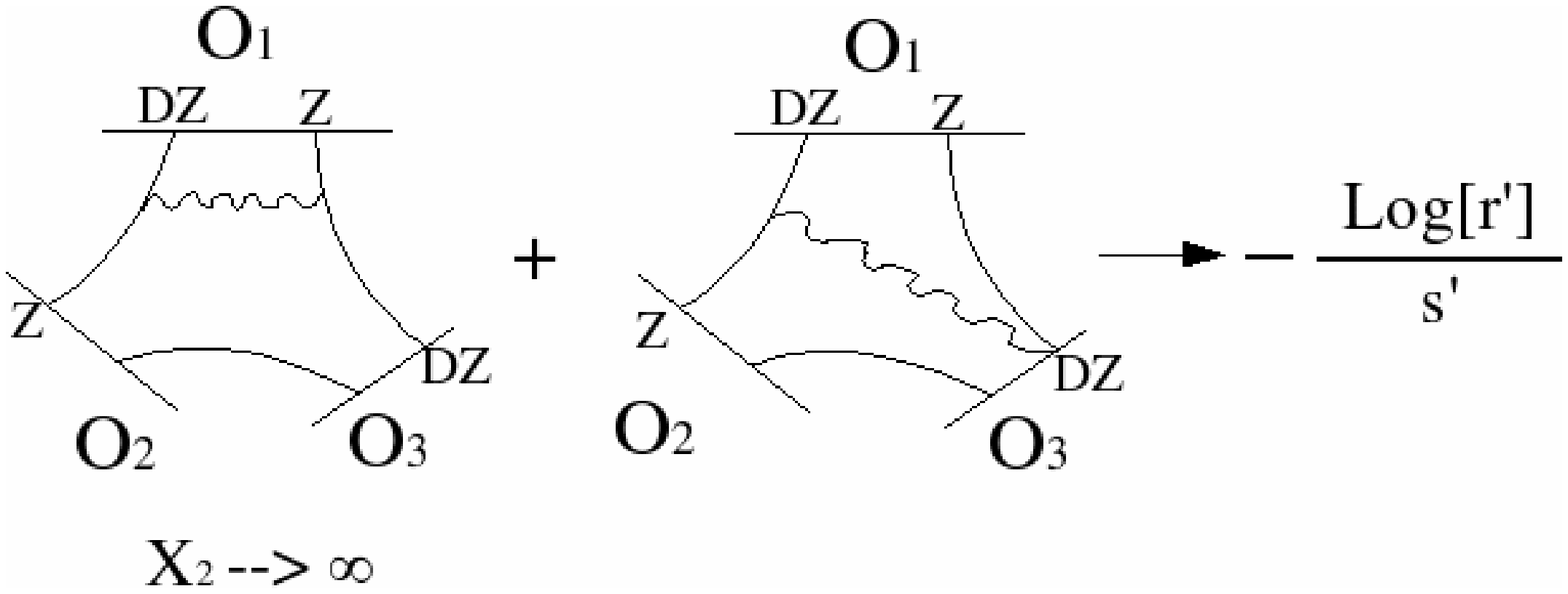}}
\caption{Differential equation ensuring conformal invariance}
}

It is now easy to generalize to the case of arbitrary $m>s; t>n$. 
For this case, the $2\rightarrow \infty$ collapse is given by
\bea{cic2}
C_2 &=& 
\delta^i_k\delta^j_l 
\frac{1}{m!n!s!t!} \partial_1^m \partial_2^n \bar\partial_3^s
\bar\partial_4^t \left( \frac{(s'-r') \phi(r',s')}{x_{13}^2 x_{24}^2
} \right), \\ \nonumber
&=& 
\delta^i_k\delta^j_l 
\frac{ {\;}^tC_n }{m!n!s!t!}
\left( (\partial_2\bar\partial_4)^n \frac{1}{x_{24}^2 } \right)
\partial_1^m \bar\partial_3^s \bar\partial_4^{t-n-1}  \times \\
\nonumber
& &\left[ \frac{1}{x_{13}^4 } ( - z_{14} \phi - z_{14} (s'-r')
\partial_{s'} \phi ) \right].
\eea
In the second line of the above equation we have first used the Leibnitz
rule 
to move the $n$ derivatives in the direction of $\bar z_4$ to act
on the term in the round bracket, then we have focussed only on the 
term which contributes to the identity $\delta_{z\bar{z}}$.  
the term in the square bracket is obtained by the action of one of
the remaining $t-n$
$\bar\partial_4$ derivatives on the collapsed term.
Now consider $A_2$, again focusing on the term which contributes to
the identity we get
\bea{cia2}
A_2 &=&
\delta^i_k\delta^j_l 
\frac{ {\;}^tC_n }{m!n!s!t!}
\left( (\partial_2\bar\partial_4)^n \frac{1}{x_{24}^2 } \right)
\partial_1^m \bar\partial_3^s \bar\partial_4^{t-n-1}  \times \\
\nonumber
& & \left[ \frac{1}{x_{13}^4} (
2 z_{13} \phi + 2 z_{13} (r'\partial_{r'} + s'\partial_{s'} ) \phi
- z_{14} \partial_{s'} \phi ) \right].
\eea
Here we have only looked at the term $p=t-n$ as it is the only one
term in the summation of  \eq{dexg4}  which
contributes to the identity. The last line in the above equation is
obtained by the action of the operator $ (2\bar\partial_3 +
\bar\partial_4)$ on $\phi(r',s')/x_{13}^2$. 
From the structure of derivatives in \eq{cic2} \eq{cia2}, it is easy
to see that only holomorphic derivatives 
acting on the term in the square brackets of these equations 
is $\partial_1$,
Therefore, for the purposes of identifying the term proportional to
the identity one can just treat the  $z's$  in these brackets as
$z_1$. Then adding \eq{cic2} and \eq{cia2}, we see that we can use
the differential equation in \eq{deaux} to obtain
\be{sumcica}
C_2 + A_2 = 
 \frac{\delta^i_k\delta^j_l }{m! s! (t-n)!x_{24}^{2(1+n)}
} 
\partial_1^m \bar\partial_3^s \bar\partial_4^{t-n-1} \left[
\frac{z_1}{x_{13}^2  x_{14}^2 } \log\left( \frac{x_{13}^2
}{x_{34}^2} \right) \right]. 
\ee
To perform the differentiation in the above equation it is
convenient to first 
do all the $\bar\partial_4$ and the $\bar\partial_3$
derivatives before finally performing the $\partial_1$ derivatives.
This gives
\bea{sumcicaf}
C_2 + A_2 &=&  
\lim_{x_2\rightarrow x_1}
\frac{\delta^i_k\delta^j_l}{(m-s) x_{24}^{2(1+n)}
x_{14}^{2(m-s)} x_{13}^{2(1+ s)} } \left( \log 
\left( \frac{x_{13}^2 }{x_{34}^2 } \right) + h(s) \right), \\
\nonumber
&=&  \frac{\delta^i_k\delta^j_l} 
{ (m-s)  x_{14}^{2(1 + t)} x_{13}^{2(1+s)}  }
\left( \log \left( \frac{x_{13}^2 }{x_{34}^2 } \right) + h(s) \right). 
\eea
Here we have also written down the final limit to be taken, note that
powers of $x$ and the presence of the log 
or the constant agrees with conformal invariance. Thus, using the
differential equation in \eq{deaux} we have shown that 
the terms $A_2$ and $C_2$ which can potentially violate conformal
invariance combine together using \eq{deaux} to restore it.
In \eq{sumcicaf} $h(s)$ refers to the harmonic number
\be{defharmo}
h(s) = \sum_{j=1}^s \frac{1}{s}, \;\;s \neq 0, 
\;\;\;\;\;\;\;\;\;\;\;\;\;\; h(0) =0.
\ee
From the tables 1. and 2.  we see that the collapse $C_3$ and the
diagram $A_3$ also contributes when $m>s$. Though these are not
dangerous diagrams one can use similar manipulations to sum these. 
This gives
\be{sumcia3}
C_3 + A_3 = \frac{\delta^i_k\delta^j_l}{ (m-s) x_{14}^{2(1+t)}
x_{13}^{2(1+s)} } \left( \log \left( \frac{x_{14}^2 }{\epsilon^2}
\right) + h(n) \right).
\ee
The total contribution from these graphs is thus obtained by adding
\eq{sumcicaf} and \eq{sumcia3}. Note that on adding these terms, the
argument of the log is precisely that of what is expected for a three
body term. 

\vspace{.5cm}
\noindent
{ {\emph Case 2. $m<s, \;\; t<n $} }

From table 1. and table 2. we see that the potentially dangerous 
diagrams are $C_1$ and $A_1$. 
This case is similar to the previous one, going through similar
manipulations we can combine these diagrams use \eq{deaux} to give
\bea{cic1a1}
C_1 + A_1 &=&
- \frac{ \delta^i_k\delta^j_l{\;}^s C_m}{m!n!s!t!} \left( 
(\partial_1\bar\partial_3 )^m \frac{1}{x_{13}^2 } \right)
\partial_2^n \bar\partial_4^t \bar\partial_3^{s-m-1} 
\left( \frac{ z_2 }{x_{24}^2 x_{23}^2 } \log 
\left( \frac{x_{34}^2 }{x_{24}^2} \right) \right),  \\ \nonumber
&=& \frac{ \delta^i_k\delta^j_l }
{(s-m) x_{13}^{2(1+m)} x_{24}^{2(1+t)} x_{23}^{2(s-m)} }
\left( \log \left( \frac{x_{24}^2}{x_{34}^2} \right) + h(t) \right).
\eea
Now taking the $x_2\rightarrow x_1$ limit one obtains
\be{cic1f}
C_1 + A_1 =
 \frac{ \delta^i_k\delta^j_l }{(s-m) x_{13}^{2 (1 +s) }x_{14}^{2( 1+t) }} 
\left( \log \left( \frac{x_{14}^2}{x_{34}^2} \right) + h(t) \right).
\ee
Again we see that the terms which can possibly violate conformal
invariance add up together to restore conformal invariance.
The diagrams $C_4$ and $A_4$ for this case can also be combined using
similar manipulations to give
\be{cicia4}
C_4 + A_4 =
 \frac{ \delta^i_k\delta^j_l }{(s-m) x_{13}^{2 (1 +s) }x_{14}^{2( 1+t) }} 
\left( \log \left( \frac{x_{13}^2}{\epsilon^2} \right) + h(m) \right).
\ee

\vspace{.5cm}
\noindent
{ \emph { Case 3. $m =s, \;\; n=t$ }}

From table 1. and table 2. we see that for this case the only
diagrams that are potentially dangerous are $C_1$ and $C_2$. The
mechanisms of how these diagrams are removed is similar to the one
for the $SO(6)$ sector discussed in section 2.2. 
The sum of all the dangerous collapses among  the 
three terms in \eq{dbasic} cancel among each other. 
For notational convenience we choose $m_a =m, m_{a+1} = n, n_{b+1} =
s, s_c =t$ in \eq{dbasic}. Then if the first term has to contribute,
we must have $n_b = s_{c+1} =0$. This is because the operator
$O_\beta$ and $O_\gamma$ have only anti-holomorphic derivatives and
the only way  
the last free contraction can contribute 
to the term proportional to the identity is when there are 
no derivatives present on the corresponding letters. 
The $SO(6)$ structure of all the three terms
involving the 
dangerous collapses \eq{dbasic} is identical so for 
convenience we suppress them. 
The dangerous terms from the first term in \eq{dbasic}
are given by
\bea{ddc1}
D(1;34) 
&=& \lim_{x_2 \rightarrow x_1} \frac{1}{(m!)^2 (s!)^2 } 
\frac{1}{x_{34}^2 }\times \\
\nonumber
& &\left[ 
(\partial_1\bar\partial_3)^m \left( \frac{1}{x_{13}^2 }
\right) ( \partial_2\bar\partial_4)^n 
\left( \frac{(s'-r')\phi(r',s')}{x_{24}^2} \right) \;\;\;
{\rm with}\;\; r' = \frac{x_{34}^2 }{x_{24}^2 }, \;\; s' =
\frac{x_{23}^2}{x_{24}^2 }  \right. \\ \nonumber
&+&
\left. (\partial_2\bar\partial_4)^n \left( \frac{1}{x_{24}^2 }
\right) ( \partial_1\bar\partial_3)^m 
\left( \frac{(s'-r')\phi(r',s')}{x_{13}^2} \right) \;\;\;
{\rm with}\;\; r' = \frac{x_{34}^2 }{x_{13}^2 }, \;\; s' =
\frac{x_{14}^2}{x_{13}^2 }  \right]. 
\eea
Note that in the above equation 
we have arranged the derivatives so that it contains the term
proportional to the identity.
Similarly the dangerous terms from the second term in \eq{dbasic} are
given by
\bea{dcc2}
D(3;41)  
&=& \lim_{x_2 \rightarrow x_3} \frac{1}{(m!)^2 (s!)^2 }
(\partial_1\bar\partial_4)^n \left( \frac{1}{x_{14}^2 } \right)
\times \\ \nonumber
& & \left[
( \partial_1 \bar\partial_3)^m \left ( 
\frac{(s'-r') \phi(r',s')}{x_{13}^2 x_{24}^2 } \right)
\;\;\;
{\rm with} \;\; r' = \frac{x_{14}^2}{x_{13}^2 }, \;\; s' =
\frac{x_{34}^2 }{x_{13}^2 } \right. \\ \nonumber
&+& 
\left. 
( \partial_1 \bar\partial_3)^m \left ( \frac{1}{x_{13}^2 }\right)
\left( \frac{(s'-r') \phi(r',s')}{x_{24}^2 } \right)
\;\;\;
{\rm with} \;\; r' = \frac{x_{14}^2 }{x_{24}^2 }, \;\;
s' = \frac{x_{12}^2}{ x_{24}^2} \right].
\eea
Note that on taking the respective limits 
we see that the first term of \eq{dcc2} cancels the second term of
\eq{ddc1} as $\phi(r,s)$ is a symmetric function in $r$ and $s$.
Finally the dangerous terms from the last term of \eq{dbasic} is
given by
\bea{dcc3}
D(4;13) &=& C_2 + A_2 \\ \nonumber
&=& \lim_{x_2 \rightarrow x_4} \frac{1}{(m!)^2 (s!)^2 }
(\partial_1\bar\partial_3)^m \left( \frac{1}{x_{13}^2 } \right)
\times \\ \nonumber
& & \left[
( \partial_1 \bar\partial_4)^n \left ( 
\frac{(s'-r') \phi(r',s')}{x_{23}^2 x_{14}^2 } \right)
\;\;\;
{\rm with} \;\; r' = \frac{x_{13}^2}{x_{14}^2 }, \;\; s' =
\frac{x_{34}^2 }{x_{14}^2 } \right. \\ \nonumber
&+& 
\left. 
( \partial_1 \bar\partial_4)^n \left ( \frac{1}{x_{14}^2 }\right)
\left( \frac{(s'-r') \phi(r',s')}{x_{24}^2 } \right)
\;\;\;
{\rm with} \;\; r' = \frac{x_{13}^2 }{x_{23}^2 }, \;\;
s' = \frac{x_{12}^2}{ x_{23}^2} \right].
\eea
It is now clear that on taking the limits in \eq{ddc1}, \eq{dcc2} and 
\eq{dcc3} the sum vanishes due to pair wise cancellations.
\be{sumddc}
D(1;34) + D(3;41) + D(4;13) =0.
\ee
Thus the dangerous collapses completely cancel restoring conformal
invariance. We have show this cancellations schematically in the 
\fig{fulcan}
\FIGURE{
\label{fulcan}
\centerline{\epsfxsize=16.truecm \epsfbox{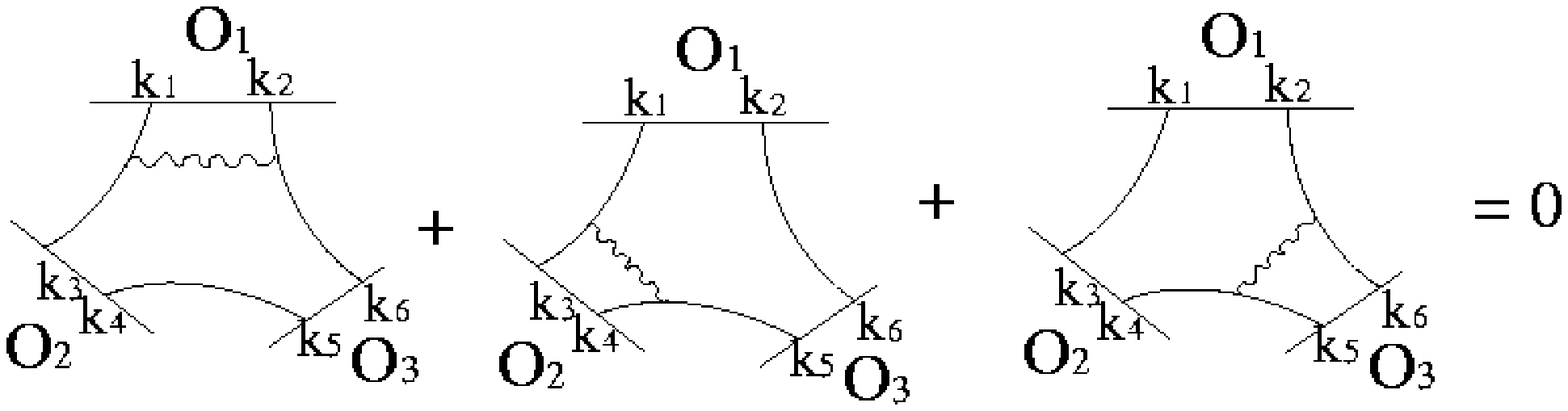}}
\caption{Cancellations among dangerous collapses}
}

From table 1. and table 2. we see that for this case of $m=s$ and
$n=t$ the collapse diagrams $C_3$ and  $C_4$ also contribute. These
diagrams are not dangerous. They are given by
\bea{c3c4}
C_3 +C_4 &=& \lim_{x_2 \rightarrow x_1} \frac{
\delta^i_k\delta^j_l }{(m!)^2 (n!)^2 } \times \\ \nonumber
& & 
\left[ (\partial_1\bar\partial_3)^m \left( \frac{1}{x_{13}^2 } \right)
(\partial_2 \bar\partial_4)^n 
\left( \frac{ (s'-r') \phi(r',s') } {x_{24}^2} \right) \right. \;\;\;{\rm with} \;\;
r' = \frac{x_{12}^2 } {x_{24}^2 }, \;\; s' = \frac{x_{14}^2}{x_{24}^2
} \\ \nonumber
& & 
\left.  (\partial_2\bar\partial_4)^n \left( \frac{1}{x_{24}^2 } \right)
(\partial_1 \bar\partial_3)^m 
\left( \frac{  (s'-r') \phi(r',s') } {x_{13}^2} \right) \right] \;\;\;{\rm with} \;\;
r' = \frac{x_{12}^2 } {x_{13}^2 }, \;\; s' = \frac{x_{23}^2}{x_{13}^2
} 
\eea
We can extract the log term and the constant by performing the
required differentiations and focusing on the contributions to the
identity. For the diagram $C_3$ and $C_4$,  
we do not need to keep track of the
constants. The reason is due to a similar phenomenon discussed for the 
$SO(6)$ sector. To obtain the renormalization group independent
constant one needs to subtract the constants from the corresponding
two body term. But, for the two body terms all the collapses
$C_1, C_2, C_3, C_4$ contribute. To find these we just write
the diagrams $C_1$ as in \eq{c1} and further take the
$x_4\rightarrow x_3$ limit. It is then easily seen that the constants
from $C_1$ is identical to the constants from $C_3$ and the constants
from $C_2$ is identical to the constants from $C_4$. 
Therefore in the renormalization group independent contribution
\be{colrgin}
C_3 (3 \rm{pt}) + C_4 ( 3\rm{pt}) 
- \frac{1}{2} \left(
C_1 (2 \rm{pt}) + C_4 ( 2\rm{pt}) +
C_3 (2 \rm{pt}) + C_4 ( 2\rm{pt})\right), 
\ee
one finds that the constants cancel. Thus we write  just the log terms
of \eq{c3c4} which contribute to the identity, these are given by
\be{c3c4l}
C_3 + C_4 = \frac{\delta^i_k\delta^j_l} { x_{13}^{2(m+1)}
x_{14}^{2(n+1)} } \left[ h(m+1) \log\left( \frac{x_{13}^2}{\epsilon^2}
\right) + h(n+1) \log\left( \frac{x_{14}^2 }{\epsilon^2} \right)
\right].
\ee

Though we have not 
emphasized length conserving processes in this paper, we mention that 
the above mechanism of ensuring conformal invariance for the
case of $m=s, n=t$ will not hold for such processes. 
For a length conserving process,  
if $O_\alpha$ is the longest operator, then there is only the first
term of \eq{sumddc}, therefore there can be no possibility of 
cancellation of the dangerous collapses. But, as we have discussed for
the case of the $SO(6)$ sector, there are non nearest neighbour
interactions which ensure cancellations of the dangerous collapses.
This is shown schematically in \fig{lccan}
\FIGURE{
\label{lccan}
\centerline{\epsfxsize=16.truecm \epsfbox{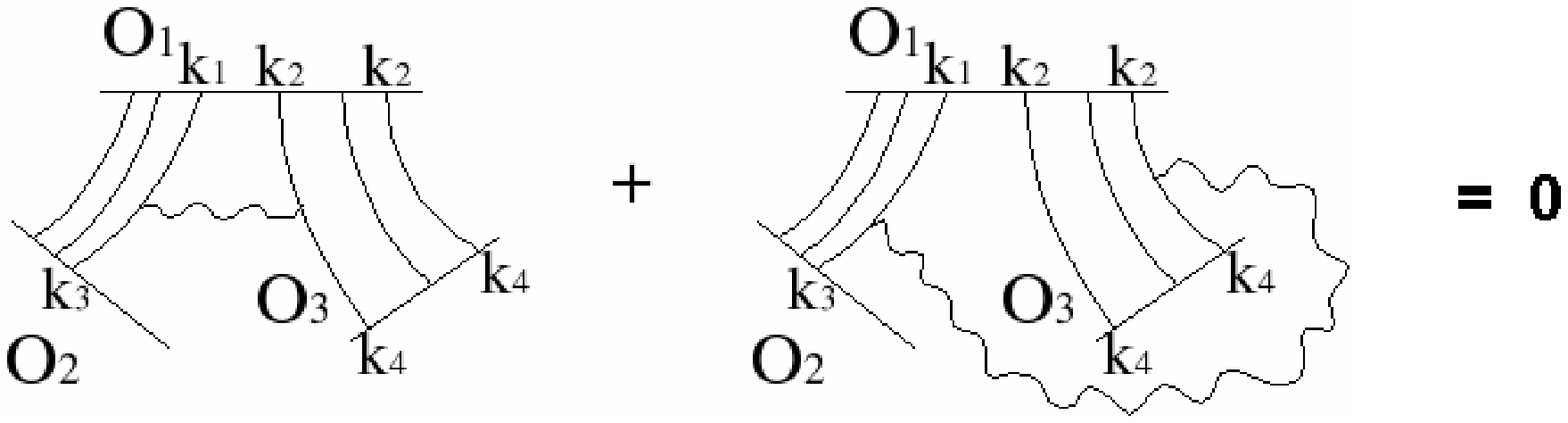}}
\caption{Cancellations in a length conserving process}
}

\subsection{Summary of the calculation} 

Here we summarize the results of our 
discussion in the previous subsections  
to give a recipe for the evaluation of one loop
corrections to structure constants for the class of operators 
with derivatives we are dealing with.
We will give the recipe to evaluate 
the constants in 
$U( 3{\rm pt}) -\frac{1}{2} U(2{\rm pt} ) $
for the various cases we have discussed.

\vspace{.5cm}
\noindent
{ \emph{ (i) Case 1. $m>s, \;\; t>n$ }}

For this case the renormalization group invariant correction to
structure constant is given by
\bea{sumfin1}
U ^{(i, m)(j,n)}_{(ks)(lt)} ( 3{\rm pt})
&-&\frac{1}{2} U ^{(i, m)(j,n)}_{(ks)(lt)} ( 2{\rm pt}) \\ \nonumber
&=& \frac{1}{2} \left( 
V^{ij}_{kl} {\cal C}_Q  + \delta^i_k\delta^j_l \left( 
{\cal C}_E  
+ C_2 + A_2 + C_3 + A_3 + B_1 \right) \right), \\ \nonumber
&=& 
\frac{1}{2} \frac{\lambda}{N}
\left( V^{ij}_{kl} {\cal C}_Q + \delta^i_k\delta^j_l \left( 
{\cal C}_E  +  \frac{h(s)}{m-s} + \frac{h(n)}{m-s} - \frac{2}{(m-s)^2
} \right)\right).
\eea
Here ${\cal C}_Q$ refers to the constant from the quartic diagram,
which can be read out from table 3. of appendix C. ${\cal C}_E$ refers
to the constant from the diagram E, this can be read out from 
the tables 4. and 5. $V^{ij}_{kl}$ stands for the $SO(6)$ structure of
the quartic given by
\be{defvijkl}
V^{ij}_{kl} = 2\delta^j_k\delta^i_l - \delta^i_k\delta^j_l -
\delta^{ij} \delta_{kl} 
\ee
In the last line of \eq{sumfin1} we have substituted the values 
constants of the diagrams  
$C_2+A_2$, $C_3+A_3$ and $B_1$ from \eq{sumcica}, \eq{sumcia3} and 
\eq{d2g1c} respectively. We have also reinstated the t'Hooft coupling
and the $1/N$ factor of the normalization of the structure constant.

\vspace{.5cm}
\noindent
{ \emph{ (ii) Case 1. $m<s, \;\; t<n$ }}

The renormalization group invariant correction to the structure
constant is given by
\bea{sumfin2}
U ^{(i, m)(j,n)}_{(ks)(lt)} ( 3{\rm pt})
&-&\frac{1}{2} U ^{(i, m)(j,n)}_{(ks)(lt)} ( 2{\rm pt}) \\ \nonumber
&=& \frac{1}{2} \left( 
V^{ij}_{kl} {\cal C}_Q  + \delta^i_k\delta^j_l \left( 
{\cal C}_E  
+ C_1 + A_1 + C_4 + A_4 + B_2 \right) \right), \\ \nonumber
&=& 
\frac{1}{2} \frac{\lambda}{N} 
\left( V^{ij}_{kl} {\cal C}_Q + \delta^i_k\delta^j_l \left( 
{\cal C}_E  +  \frac{h(t)}{s-m} + \frac{h(m)}{s-m} - \frac{2}{(m-s)^2
} \right)\right).
\eea
Here we have substituted the values of $C_1 +A_1$, $C_4 +A_4$ and
$B_2$ from \eq{cic1f}, \eq{cicia4} and \eq{d2g22}. The rest of the
constants can be read out from the tables in appendix C.

\vspace{.5cm}
\noindent
{ \emph{ (iii) Case 2. $m=s, \;\; t=n$ }}

As we have discussed earlier for this case the constants from all the
collapses cancel in the renormalization group invariant combination 
given in \eq{colrgin}. There are no contributions from gauge bosons on
two external legs, 
thus we are left with constants only from the 
quartic $Q$ and the diagram $E$, therefore we have
\bea{sumfin3}
U ^{(i, m)(j,n)}_{(ks)(lt)} ( 3{\rm pt})
&-&\frac{1}{2} U ^{(i, m)(j,n)}_{(ks)(lt)} ( 2{\rm pt}) \\ \nonumber
&=& \frac{1}{2} \frac{\lambda}{N} \left( 
V^{ij}_{kl} {\cal C}_Q  + \delta^i_k\delta^j_l \left( 
{\cal C}_E  \right) \right).
\eea
Again the constants occurring above can be read out from appendix C. 
As a simple check note that when the number of derivatives are set to
zero, evaluating ${\cal C}_Q$ and ${\cal C}_E$ in the above we obtain 
the anomalous dimension Hamiltonian ${\cal H}$ which determines the 
corrections to structure constants in the $SO(6)$ sector.

\subsection{An example}

To illustrate the methods developed we compute the one loop
corrections for  
a simple example of three point function.
Consider the following three operators:
\bea{defderop}
O_\alpha &=& \frac{1}{\sqrt{N^3} } \sum_{k=0}^n {\;}^nC_k (-1)^k
{\rm Tr}  ( \partial^{n-k} \phi^1 \partial^k \phi^2 \phi^3 ),
\\ \nonumber
O_\beta &=& \frac{1}{\sqrt{N^3} } \sum_{k=0}^n {\;}^nC_k (-1)^k
{\rm Tr}  ( \bar\partial^{n-k} \phi^1 \bar\partial^k \phi^2 \phi^4 ),
\\ \nonumber
O_\gamma &=& \frac{1}{N} {\rm Tr} ( \phi^3 \phi^4).
\eea
where $O_\alpha$ is at position $x_1$, $O_\beta$ at $x_3$ and
$O_\gamma$ at $x_4$. 
The tree level correlation function of these operators is given by
\be{treeder}
\langle O_\alpha O_\beta O_\gamma \rangle ^{(0)} =
\frac{1}{N} \sum_{k=0}^n \frac{ ({\;}^n C_k)^2 }{x_{13}^{2(n+1)} x_{14}^2
x_{34}^2} 
\ee

Now we compute the one loop corrections to this structure constant. 
All the corrections, the log terms as well as the renormalization
group invariant correction will multiply the 
position dependent prefactor
\be{ppred}
\frac{1}{{x_{13}^{2(n+1)} x_{14}^2 x_{34}^2}},
\ee
which is determined by the tree level dimensions of the three
operators in \eq{defderop}. 
We write below the log corrections and the renormalization group
invariant correction to the structure constant arising from the
various diagrams.

\vspace{.5cm}
\noindent
{ \emph {Three body terms} }

The three body interactions consists of the following diagrams:
\bea{f3body}
& & 2 \sum_{k=0}^n ({\;}^nC_k)^2
\left( Q^{k0}_{k0} + E^{k0}_{k0} + (C_3 + C_4)^{k0}_{k0} (1;34)
\right. \\ \nonumber
&+& \left.
Q^{k0}_{k0} + E^{k0}_{k0} + (C_3 + C_4)^{k0}_{k0} (3;41)
+ (C_3 + C_4)^{00}_{00} (4;13)  \right).
\eea
Here we have suppressed the $SO(6)$ indices but kept the indices
which indicate the number of derivatives on the letters involved. 
There are no contributions of $(Q + E) (4;13)$ as the $SO(6)$ 
structure of these diagrams ensures that they cancel each other. 
Evaluating the log terms of these diagrams  using the tables in appendix
C. we find: 
\bea{f3logbody}
\nonumber
& &2 \sum_{k=0}^n   {\;}^n C_k)^2  \left(
 \left[ -\frac{2}{k+1} - h(k) \right]
\log\left( \frac{x_{13}^2 x_{14}^2 }{ x_{34}^2 \epsilon^2 } \right)  
+ h(k+1) \log \left( \frac{x_{13}^2} {\epsilon^2 } \right)
+ \log \left( \frac{x_{14}^2 }{\epsilon^2 } \right) 
\right. \\
\nonumber
&+&
\left[ -\frac{2}{k+1} - h(k) \right] \log 
\left( \frac{x_{13}^2 x_{34}^2 }{ x_{14}^2 \epsilon^2 } \right) 
+ h(k+1) \log \left( \frac{x_{13}^2} {\epsilon^2 } \right)
+ \log \left( \frac{x_{34}^2 }{\epsilon^2 } \right) \\ 
&+& \left. \log\left( \frac{x_{14}^2}{\epsilon^2 }\right) + 
 \log\left(\frac{x_{34}^2}{\epsilon^2} \right) \right).
\eea
We have written down  each contribution in \eq{f3logbody}, so that
they appear in the order of the diagrams in \eq{f3body}.
To write the renormalization group invariant correction to the 
structure constants we need to find the constant in each of the terms
in \eq{f3body} and perform the metric subtractions. We
have already shown that the  constants form all the collapses 
in \eq{f3body} cancel. Therefore we have to look for constants of only the 
$Q$'s and $E$'s which are listed in appendix C. The metric
contributions to these are identical and since they are
weighted by $1/2$, the final result is just half of the corresponding
values listed in appendix C. Writing down these for each of the terms
in \eq{f3body} we get
\be{conresu}
{\cal K} = - 4\sum_{k=0}^n ({\;}^nC_k)^2  
\times\left( 
\sum_{l=0}^{k} 
 (-1)^l {\;}^kC_l \frac{l+k+2}{(l+1)^2}  h(l+1) \right). 
\ee
Note that if the number of derivatives $n$ is set to zero in the above
expression we obtain $-8$ which agrees with \eq{strcon}.

\vspace{.5cm}
\noindent
{\emph{ Two body terms}}

As we have discussed before, because of the slicing argument one needs
to evaluate only the  terms proportional to the 
logarithm in the two body diagrams. The diagrams
are given by
\bea{d3blog}
& \sum_{k, k' =0}^n {\;}^nC_k {\;}^nC_k' 
(-1)^{k+k'} 
 \left( Q + E  \right. \\ \nonumber   
&+ \left. C_1 + C_2 + C_3 + C_4 + A_1 + A_2 + A_3 + A_4
\right)^{k n-k}_{k' n-k'} ( 1;3 ) \\ \nonumber
&+ \sum_{k=0}^n ({\;}^n C_k )^2 
 \left( S_k( 1; 3) + S_{n-k} (1;3) + S_0(1;4) + S_0(3;4) \right),
\eea
where $S_k$ refers to the self energy contribution of a scalar with 
$k$ derivatives. The contribution of these 
self energy diagrams can be read out from
\cite{Beisert:2003jj}.
Evaluating the  terms proportional to the logarithm 
of these diagrams we obtain
\bea{and3blog}
& &\sum_{k=0}^n ({\;}^nC_k)^2 \left(
( -2 h(k)  - \frac{2}{n+1} ) 
\log\left( \frac{x_{13}^4}{ \epsilon^4 } \right)
+   4 h(k+1)   \log\left( \frac{x_{13}^2}{\epsilon^2}
\right) \right) \\ \nonumber
&+& \sum_{k, k', k\neq k'}^n   
{\;}^nC_k {\;}^n C_{k'} (-1)^{k+k'}
\left(
( \frac{1}{|k-k'|} - \frac{2}{n+1} ) 
\log\left( \frac{x_{13}^4}{ \epsilon^4 } \right) 
+ \frac{2}{|k-k'| } 
\log \left( \frac{x_{13}^2}{\epsilon^2} \right) \right)
\\ \nonumber
&-& 4 \sum_{k=0}^n ({\;}^n C_k)^2  \left[
\left(  h(k) + h(k+1)    +1 \right) 
\log\left( \frac{x_{13}^2 }{\epsilon^2} \right) 
+  \log\left( \frac{x_{14}^2}{\epsilon^2} \right)
+  \log\left( \frac{x_{34}^2}{\epsilon^2} \right).
\right]
\eea
Adding the log terms in \eq{f3logbody} and \eq{and3blog} we obtain
only terms with  $\log( { {x_{13}^2}}/ {\epsilon^2} )$. The rest of the
log terms cancel, this coefficient is given by:
\bea{fdlogan}
& &-4\sum_{k=0}^n ({\;}^nC_k)^2 
\left(
 \frac{1}{ k+1}  +2 h(k)  +1 \right) 
 - 4 \delta_{n,0} 
 \\ \nonumber
&+& \sum_{k, k', k\neq k' }^n {\;}^nC_k {\;}^n C_{k'} (-1)^{k+k'} \left( 
\frac{4}{|k-k'|} \right).
\eea
As a simple check, note that on setting $n=0$ the above expression
reduces to $-12$ which was obtained in \eq{allscalcor}.

\section{Conclusions}

We have evaluated one loop corrections to the structure constants in
planar ${\cal N}=4$ Yang-Mills for two classes of operators, the 
$SO(6)$ sector and for operators with derivatives in 
one holomorphic direction. The summary of the results which enables
one to evaluate these structure constants for any operator in these
sectors are given in section 4.4. For the $SO(6)$  
scalar sector we find that the one loop
anomalous dimension Hamiltonian determines the corrections to the
structure constants. The reasons for this are: ${\cal N}=4$
supersymmetry which relates the quartic coupling of scalars to the 
gauge coupling, the $SO(6)$ spin dependent term factorizes in the 
calculations and contributions of all the collapsed diagrams
canceled. 
For the sector with derivatives we noticed that essentially the
structure  constants are determined by a suitable combination of derivatives
acting on the fundamental tree function $\phi(r,s)$. 
Conformal invariance in the calculation was ensured by a linear
inhomogeneous partial differential equation satisfied by $\phi(r,s)$
which enabled us to combine the diagrams violating conformal
invariance to restore it. The methods developed in this paper
can be generalized to the all classes of operators in ${\cal N}=4$
Yang-Mills.

The fact that in the $SO(6)$ sector the one loop corrections to
the structure constants are determined by the 
one loop anomalous dimension
Hamiltonian indicates the possibility that 
in a string bit theory the one loop corrected structure constants
can be determined by the delta function overlap with
modification in the propagation
of the bits taken into account.
The immediate suggestion would be that it is the anomalous dimension 
Hamiltonian which determines the propagation of the bits. 
In \cite{adgn:2005} we address this question in detail.

\acknowledgments

J.R.D would like to thank the 
discussions and  hospitality at CERN;
Harish Chandra Research Institute,
Allahabad; Institute of Mathematical Sciences, Chennai  and 
Tata Institute of Fundamental research, Mumbai; during the course of
this project. We thank the organizers of the 
Indian strings meeting, 2004 at  Khajuraho for the opportunity to present
this work. We would like to thank Avinash Dhar in particular for
stimulating discussions and criticisms
The work of the authors is partially supported by the RTN European
program: MRTN-CT-2004-503369.

\appendix

\section{Notations}

The action of ${\cal N}=4$
supersymmetric Yang-Mills is best thought of as dimensional reduced
maximal supersymmetric Yang-Mills from 10 dimensions. The action is
given by
\bea{ymaction}
S = \frac{1}{(2\pi)^2 } \int
d^4 x {\rm Tr} \left( 
\frac{1}{4} F_{\mu\nu}^{\mu\nu} + 
\frac{1}{2} D_\mu \phi^i D^\mu \phi^i
-\frac{g^2 }{4} [\phi^i, \phi^j][\phi^i, \phi^j ]  
\right. \\ \nonumber
\left. + \frac{1}{2} \bar{\psi} \Gamma_\mu D^\mu \psi 
- g\frac{i}{2} \bar{\psi} \Gamma_i[\phi^i,\psi] \right),
\eea
where $A_\mu$ with  $\mu = 1, \ldots, 4$  is the gauge field in 
4 dimensions, $\psi$ is a 16 component Majorana-Weyl spinor obtained
from the Majorana-Weyl spinor in 10 dimensions. $\phi^i$, $i =1,\ldots
6$ are scalars which transform as a vector under the R-symmetry group
$SO(6)$. $(\Gamma_\mu, \Gamma_i)$ are the ten-dimensional Dirac
matrices in the Majorana-Weyl representation.
All the fields transform in the adjoint representation of 
the gauge group $U(N)$, to be specific they are $N\times N$ matrices
which can be expanded in terms of the generators $T^a$ of the gauge 
group as
\be{gexpfie}
\phi^i = \sum_{a=0}^{N^2 -1} \phi^{i (a)} T^a, \;\;\;\;
A_\mu = \sum_{a=0}^{N^2 -1} A_\mu^{ (a)} T^a, \;\;\;\;
\psi = \sum_{a=0}^{N^2 -1} \psi^{ (a)} T^a. 
\ee
The generators $T^a$ satisfy
\be{condgen}
{\rm Tr}(T^a T^b) = \delta^{ab}, 
\;\;\;\;\; \sum_{a=0}^{N^2-1}
(T^a)^\alpha_\beta (T^a)^\gamma_\delta = \delta^\alpha_\delta
\delta^\gamma_\beta.
\ee
In \eq{ymaction} $g^2 = g_{YM}^2 /2(2\pi)^2$, 
\footnote{Our
convention differs from 
\cite{Beisert:2002bb} in that we have scaled the fields by
$g_{YM}/2\pi\sqrt{2} $}
the covariant derivatives are given by $D_\mu = \partial_\mu + i
g [A_\mu, \;\cdot\;] $, and $F_{\mu\nu} = \partial_\mu A_\nu
-\partial_\nu A_\mu + i g^2 [A_\mu , A_\nu]$. All our
calculations are done in the Feynman gauge. 
Using the normalization of the action given in \eq{ymaction}, the 
tree level two point functions 
of the scalar and the vector  are given by
\bea{tree2ptact}
\langle \phi^{i(a)}(x_1) \phi^{j(b)} (x_2)\rangle &=&
\frac{\delta^{ij}
\delta^{ab}}
{(x_1 -x_2)^2}, \\ \nonumber
\langle A_\mu^{(a)}(x_1) A_\nu^{(b)}(x_2) \rangle &=&
\frac{\delta_{\mu\nu}
\delta^{ab}}
{(x_1 -x_2)^2}. 
\eea

\section{Properties of the fundamental tree function}

In this appendix we will prove various properties of the
fundamental tree function $\phi(r,s)$ defined in \eq{defiphi} which
are used at various instances in the paper.
To obtain a series expansion of $\phi(r,s)$ and to show that it
satisfies the partial differential equation \eq{deaux} 
we will use is its integral
representation shown in \cite{Usyukina:1992jd} 
\be{intrephi}
\phi(r,s)=\int_0^1 \frac{-\log{(r/s)}-2\log \xi}{s-\xi (r+s-1)+
\xi^2 r }~d\xi .
\ee
From this integral representation we can find a series expansion of
$\phi(r,s)$ around $r=0,s=1$, by 
expanding the denominator in \eq{intrephi} as
\be{denexp}
\frac{1}{s-\xi (r+s-1)+ \xi^2 r }=\sum_{k,l=0}^{\infty} (-1)^{k+l}
\xi^k (\xi-1)^{k+l} \frac{(k+l)!}{k!~l!}r^k (1-s)^l.
\ee
To perform the series expansion
we need the following integrals
\bea{useinte}
\int_0^1 \xi^k (\xi-1)^{k+l}~ d \xi &=& (-1)^{k+l} \frac{k!
(k+l)!}{(2k+l+1)!},\\ \nonumber
\int_0^1 \xi^k (\xi-1)^{k+l}\log \xi ~ d \xi &=& (-1)^{k+l} \frac{k!
(k+l)!}{(2k+l+1)!}\left(h(k)-h(2k+l+1)\right),
\eea
where $h(n)$ is the harmonic number defined in \eq{defharmo}. Substituting
\eq{useinte} and \eq{denexp} in \eq{intrephi} we obtain
\begin{eqnarray}
\phi(r,s) &=& -\sum_{k,l=0}^{\infty} \frac{(k+l)!^2}{l! (2k+l+1)!}r^k
(1-s)^l \log{(r/s)}\\ \nonumber
&+&2\sum_{k,l=0}^{\infty} \frac{(k+l)!^2}{l!
(2k+l+1)!}\left(h(2k+l+1)-h(k)\right)r^k (1-s)^l.
\end{eqnarray}
Through out the paper we need the expansion of $\phi(r,s)$ at $r=0$,
this is given by
\bea{usexp}
\phi(0,s) &=& - \sum_{l=0}^\infty \frac{1}{l+1} ( 1-s)^l \ln (\frac{r}{s}
) + 2 \sum_{l=0}^\infty h(l+1) \frac{1}{l+1} (1-s)^l , \\ \nonumber
& =& - \sum_{l=0}^\infty \frac{1}{l+1} (1-s)^l  \ln(r) + 2
\frac{(1-s)^l}{(l+1)^2}
\eea

Now we show that 
$\phi(r,s)$ satisfies the following inhomogeneous linear partial
differential equations which ensures conformal
invariance in the three point function calculations of the paper.
\bea{de}
\phi(r,s)+(s+r-1)\partial_s \phi(r,s)+2 r \partial_r
\phi(r,s)=-\frac{\log{r}}{s},\\
\phi(r,s)+(s+r-1)\partial_r \phi(r,s)+2 s \partial_s
\phi(r,s)=-\frac{\log{s}}{r}.
\eea
To, simplify matters, we  introduce  the notation
\be{defnota}
D(r,s,\xi)=s-\xi(r+s-1)+\xi^2 r, 
\ee
then substituting the integral representation 
\eq{intrephi} of $\phi(r,s)$ in 
the first equation of \eq{de} we obtain 
\begin{eqnarray}
& &(1+(s+r-1)\partial_s +2 r
\partial_r)\phi(r,s) = \\ \nonumber
& &
\int^1_0 d\xi \frac{1}{D(r,s,\xi)}
\left(-\log{r/s}-2\log{\xi}+(s+r-1)/s-2\right) 
\\ \nonumber
&+&\int_0^1 d\xi
\frac{\log{r/s}+2\log{\xi}}{(D(r,s,\xi))^2}((s+r-1)\partial_s
D(r,s,\xi)+2r \partial_r D(r,s,\xi)) .
\end{eqnarray}
We can integrate the expression on the second line of the above
equation by parts by
using the following identity
\begin{equation}
(s+r-1)\partial_s D(r,s,\xi)+2r \partial_r
D(r,s,\xi)=-(1-\xi)\partial_\xi D(r,s,\xi).
\end{equation}
which results in 
\begin{eqnarray}
\label{splitdiv}
(1+(s+r-1)\partial_s +2 r
\partial_r)\phi(r,s)&=& \left. \frac{(1-\xi)(\log{r/s}+2\log{\xi}))}{D(r,s,\xi)}
\right|^1_\epsilon+ \nonumber\\
&+&\int_\epsilon^1 d\xi \frac{(s+r-1)/s-2/\xi)}{D(r,s,\xi)}
\end{eqnarray}
Note that we have introduced and parameter $\epsilon$ since $\log \xi$
in the first term is divergent at the lower limit. Similarly there is 
a log divergence in the second term of the above equation.
We now show that these divergences cancel each other. 
Let us write the term contributing to the divergence in the second
term of \eq{splitdiv} as
\begin{eqnarray}
\int_\epsilon^1 d\xi \frac{-2/\xi}{D(r,s,\xi)}=\int^1_\epsilon d\xi
\frac{-2/s}{\xi}+\int^1_\epsilon d\xi \frac{-2(r+s-1 -r
\xi)/s}{D(r,s,\xi)}
\end{eqnarray}
Substituting this in \eq{splitdiv} we obtain
\begin{eqnarray}
(1+(s+r-1)\partial_s +2 r
\partial_r)\phi(r,s)&=&
\left. \frac{\log{r/s}-\xi(\log{r/s}+2\log{\xi})}{D(r,s,\xi)}\right|^1_0
\nonumber\\
&+&\int^1_0\frac{(-(r+s-1)+2r \xi)/s}{D(r,s,\xi)},\\ \nonumber
&=&-\left. \frac{\log{r/s}}{s}+\frac{\log{D(r,s,\xi)}}{s}\right|^1_0,
\\ \nonumber
&=&-\frac{\log{r}}{s}.
\end{eqnarray}
Using similar manipulations one can show that $\phi(r,s)$ also
satisfies the second partial differential equation in \eq{de}.

We also use the fact that $\phi(r,s)$ is a symmetric function in $r$
and $s$. This is best shown using the defining expression of 
$\phi(r,s)$
\be{definephi}
\phi(r,s) = \frac{x_{13}^2 x_{24}^2}{\pi^2} \int 
d^4 u \frac{1}
{ (x_1 -u)^2  (x_2 -u)^2  (x_3 -u)^2 (x_4 -u)^2 },
\ee
where $r$ and $s$ are given by
\be{apdefrs}
r = \frac{x_{12}^2 x_{34}^2}{x_{13}^2 x_{24}^2 } , \;\;\;\;\;\;
s = \frac{x_{14}^2 x_{23}^2}{x_{13}^2 x_{24}^2 }. 
\ee
From the definition of $r$ and $s$ above we see that interchange of
$x_1$ and $x_3$ brings about an interchange of $r$ and $s$. 
But the definition \eq{definephi} is easily seen to be invariant under
$x_1$ to $x_3$.  Therefore, we conclude $\phi(r,s)$ is a symmetric
function of $r$ and $s$. $\phi(r,s)$ also satisfies the property
\be{hydensym}
\phi(r,s)=\frac{1}{r}
\phi(1/r,s/r).
\ee
This can be shown from the fact $r \leftrightarrow 1/r$ and 
$s\leftrightarrow s/r$ when $x_2\leftrightarrow x_3$. 
Then it is easy to see that the symmetry \eq{hydensym} is manifest in
\eq{definephi}. Though these symmetry properties of $\phi(r,s)$ are
not manifest in its integral representation given in \eq{intrephi}, 
we have seen that through a series of manipulations it is possible
to derive these symmetry properties from \eq{intrephi}.

\newpage

\section{Tables}

In the table below we given the values of the coefficient of 
the logarithm  ${\cal A}_Q$ and the constant ${\cal C}_Q$ of the 
quartic $Q$ in \eq{genquart}.

\begin{center}
\begin{tabular}{l | l | l | l| l| l}
$m$ & $n$ & $s$ & $t$ & ${\cal A}$  & ${\cal C}$ \\
\hline
$\;$ & $\;$& $\;$& $\;$& $\;$& $\;$ \\ 
$m$ & $0$ & $m$ & 0& $\frac{1}{m+1}$  & 
$  \sum_{l=0}^m \frac{2h(l+1)}{l+1} (-1)^l {\;} ^m C_l$
\\ 
$\;$ & $\;$& $\;$& $\;$& $\;$& $\;$ \\ 
\hline
$\;$ & $\;$& $\;$& $\;$& $\;$& $\;$ \\ 
$m$ & $0$ & $0$ & $m$ & $\frac{1}{m+1}$ & $ \frac{2}{(m+1)^2}$
\\ 
$\;$ & $\;$& $\;$& $\;$& $\;$& $\;$ \\ 
\hline
$\;$ & $\;$& $\;$& $\;$& $\;$& $\;$ \\ 
 $0$  & $m$ & $m$ & $0$  & $\frac{1}{m+1}$ &  $  \frac{2}{(m+1)^2
 }$ \\
$\;$ & $\;$& $\;$& $\;$& $\;$& $\;$ \\ 
\hline
$\;$ & $\;$& $\;$& $\;$& $\;$& $\;$ \\ 
$m$ & $n$ & $s$ & $0$ & $\frac{1}{s+1}$ & $  
-\frac{h(s)}{ s+1} + {\;}^sC_m 
\sum_{l=0}^m (-1)^{m-l} {\;}^mC_l \left( \frac{ h(s-l)}{s-l+1}  +
\frac{2}{(s-l+1)^2}  \right)$ \\ 
$\;$ & $\;$& $\;$& $\;$& $\;$& $\;$ \\ 
\hline
$\;$ & $\;$& $\;$& $\;$& $\;$& $\;$ \\ 
$m$ & $n$ & $0$ & $t$ & $\frac{1}{t+1}$ & $  
-\frac{h(t)}{ t+1} + {\;}^tC_n 
\sum_{l=0}^n (-1)^{n-l} {\;}^nC_l \left( \frac{ h(t-l)}{t-l+1}  +
\frac{2}{(t-l+1)^2}  \right)$ \\ 
$\;$ & $\;$& $\;$& $\;$& $\;$& $\;$ \\ 
\hline
$\;$ & $\;$& $\;$& $\;$& $\;$& $\;$ \\ 
$m$ & $0$ & $s$ & $t$ & $\frac{1}{m+1}$ & $  
-\frac{h(m)}{ m+1} + {\;}^mC_s 
\sum_{l=0}^s (-1)^{s-l} {\;}^sC_l \left( \frac{ h(m-l)}{m-l+1}  +
\frac{2}{(m-l+1)^2}  \right)$ \\ 
$\;$ & $\;$& $\;$& $\;$& $\;$& $\;$ \\ 
\hline
$\;$ & $\;$& $\;$& $\;$& $\;$& $\;$ \\ 
$0$ & $n$ & $s$ & $t$ & $\frac{1}{m+1}$ & $  
-\frac{h(n)}{ n+1} + {\;}^nC_t 
\sum_{l=0}^t (-1)^{t-l} {\;}^tC_l \left( \frac{ h(n-l)}{n-l+1}  +
\frac{2}{(n-l+1)^2}  \right)$ \\ 
$\;$ & $\;$& $\;$& $\;$& $\;$& $\;$ \\ 
\hline
\end{tabular}

\vspace{.5cm}
{\small{\bf{ Table 3: } } } ${\cal A}_Q$ and ${\cal C}_Q$ for the quartic
$Q$.
\end{center}

Note that we have not given the values of ${\cal A}_Q$ and ${\cal C}_Q$
for the most general case of $m, n, s, t$.  
The value of the term proportional to the logarithm ${\cal A}_Q$, is 
always $1/(m+n+1)$ for arbitrary values of $m, n,s, t$. 
The manipulations to extract the constant from 
\eq{genquart} for arbitrary values of $m, n,s,t$ are considerably
more involved, but one
can in principle extract  the value of
${\cal C}_Q$ using Mathematica routines, 
we have not attempted to do so.

\newpage

In the table below we list the coefficient of 
the logarithm and the constant for the gauge exchange diagram $E$ 
of \eq{genex}.
\begin{center}
\begin{tabular}{ l | l | l | l| l| l}
$m$ & $n$ & $s$ & $t$ & ${\cal A}_E$ &  ${\cal C}_E$
\\ \hline
$\;$ & $\;$& $\;$& $\;$& $\;$ & $\;$ \\ 
$m$ & $0$ & $m$ & $0$ & $-h(m) - \frac{1}{m+1}$ & 
$ -(m+1) \sum_{l=0}^m 
\frac{2h(l+1)}{(l+1)^2}(-1)^l {\;}^mC_l  $ \\
$\;$ & $\;$& $\;$& $\;$& $\;$ & $\;$ \\ 
\hline
$\;$ & $\;$& $\;$& $\;$& $\;$ & $\;$ \\ 
$0$ & $n$ & $0$ & $n$ & $-h(n) - \frac{1}{n+1}$ & 
$ -(n+1) \sum_{l=0}^n 
\frac{2h(l+1)}{(l+1)^2}(-1)^l {\;}^nC_l  $ \\
$\;$ & $\;$& $\;$& $\;$& $\;$ & $\;$ \\ 
\hline
$\;$ & $\;$& $\;$& $\;$& $\;$ & $\;$ \\ 
$m$ & $0$ & $0$ & $m$ &  $ \frac{1}{m} - \frac{1}{m+1} $ & 
$\frac{2}{m^2} - \frac{2}{(m+1)^2}$  \\
$\;$ & $\;$& $\;$& $\;$& $\;$ & $\;$  \\ 
\hline
$\;$ & $\;$& $\;$& $\;$& $\;$ & $\;$ \\ 
$0$ & $n$ & $n$ & $0$ & $ \frac{1}{n} - \frac{1}{ n+1} $ & 
$\frac{2}{n^2} - \frac{2}{(n+1)^2}$  \\
$\;$ & $\;$& $\;$& $\;$& $\;$ & $\;$ \\ 
\hline
\end{tabular}
\\
\vspace{.5cm}
{\small{\bf{ Table 4: } } } ${\cal A}_E$ and ${\cal C}_E$ for the gauge
exchange $E$.
\end{center}

To write down the value of the gauge exchange term $E$ for the other
case, it is more convenient to consider $E+ Q$, where $Q$ is the
corresponding quartic contribution. 
Since the values of the quartic term is known from table 3.
the value of $E$ is also known. Below is the table which lists the
contribution of $E+Q$ for the remaining cases of $m$, $n$, $s$, $t$.

\begin{center}
\begin{tabular}{ l | l | l | l| l| l}
$m$ & $n$ & $s$ & $t$ & ${\cal A}$ & ${\cal C}$
\\ \hline
$\;$ & $\;$& $\;$& $\;$& $\;$ & $\;$ \\ 
$m$ & $n$ & $s$ & $0$ & $\frac{1}{s-m}$ &
$ -\frac{h(m)}{s-m} + {\;}^sC_m 
\sum_{l=0}^m (-1)^{m-l}  {\;}^m C_l \frac{1} { (s-l)^2 }$ \\
$\;$ & $\;$& $\;$& $\;$& $\;$ & $\;$ \\ 
\hline
$\;$ & $\;$& $\;$& $\;$& $\;$ & $\;$ \\ 
$m$ & $n$ & $0$ & $t$ &  $\frac{1}{t-n}$ & 
$ -\frac{h(n)}{t-n} + {\;}^tC_n 
\sum_{l=0}^n (-1)^{n-l}  {\;}^n C_l \frac{1} { (t-l)^2 }$ \\
$\;$ & $\;$& $\;$& $\;$& $\;$ & $\;$\\ 
\hline
$\;$ & $\;$& $\;$& $\;$& $\;$ & $\;$ \\ 
$m$ & $0$ & $s$ & $t$ & $\frac{1}{ m-s}$ & 
$ -\frac{h(s)}{m-s} + {\;}^mC_s 
\sum_{l=0}^s (-1)^{s-l}  {\;}^s C_l \frac{1} { (m-l)^2 }$ \\
$\;$ & $\;$& $\;$& $\;$& $\;$ & $\;$ \\ 
\hline
$\;$ & $\;$& $\;$& $\;$& $\;$ & $\;$ \\ 
$ 0$ & $n$ & $s$ & $t$ & $ \frac{1}{n-t}$ &
$ -\frac{h(t)}{n-t} + {\;}^nC_t 
\sum_{l=0}^s (-1)^{t-l}  {\;}^t C_l \frac{1} { (n-l)^2 }$ \\
$\;$ & $\;$& $\;$& $\;$& $\;$ & $\;$ \\ 
\hline
\end{tabular}
\\
\vspace{.5cm}
{\small{\bf{ Table 4: } } } ${\cal A}$ and ${\cal C}$ for  
$Q+E.$
\end{center}

If $m\neq s$
the log term for $Q+E$ for arbitrary values of $m, n, s, t$ is given by
$1/|m-s|$ and for $m=s$ it is given by $- h(m) - h(n)$.
Again we have not listed the values of ${\cal C}$  for arbitrary
values of the derivatives, but they can be in principle be obtained
from \eq{genex} using routines in Mathematica.

\bibliographystyle{utphys}
\bibliography{3pt}

\end{document}